\documentclass[11pt,a4paper]{article}

\usepackage{amssymb}
\usepackage{graphicx}
\usepackage{bm}

\usepackage{mathrsfs}
\usepackage{cite}

\begin{document}

\title{The three-loop anomalous dimension and the four-loop $\beta$-function for ${\cal N}=1$ SQED regularized by higher derivatives}

\author{I.E.Shirokov\,${}^a$, K.V.Stepanyantz\,${}^{ab}$ $\vphantom{\Big(}$
\medskip\\
{\small{\em Moscow State University, Faculty of Physics,}}\\
${}^a${\small{\em Department of Theoretical Physics,}}\\
${}^b${\small{\em Department of Quantum Theory and High Energy Physics,}}\\
\medskip
{\small{\em 119991, Moscow, Russia.}}
}

\maketitle

\begin{abstract}
For ${\cal N}=1$ SQED with $N_f$ flavors regularized by higher derivatives in the general $\xi$-gauge we calculate the three-loop anomalous dimension of the matter superfields defined in terms of the bare coupling constant and demonstrate its gauge independence. After this the four-loop $\beta$-function defined in terms of the bare coupling constant is obtained with the help of the NSVZ equation, which is valid for these renormalization group functions in all loops. Next, we calculate the three-loop anomalous dimension and the four-loop $\beta$-function defined in terms of the renormalized coupling constant for an arbitrary subtraction scheme supplementing the higher derivative regularization. Then we construct a renormalization prescription for which the results coincide with the ones in the $\overline{\mbox{DR}}$-scheme and describe all NSVZ schemes in the considered approximation. Also we demonstrate the existence of a subtraction scheme in which the anomalous dimension does not depend on $N_f$, while the $\beta$-function contains only terms of the first order in $N_f$. This scheme is obtained with the help of a finite renormalization compatible with a structure of quantum corrections and is NSVZ. The existence of such an NSVZ scheme is also proved in all loops.
\end{abstract}

\section{Introduction}
\hspace*{\parindent}

Calculations of higher order quantum corrections in ${\cal N}=1$ supersymmetric theories are important for both theory and phenomenological applications \cite{Mihaila:2013wma}. Most of these calculations (see, e.g., \cite{Avdeev:1981ew,Jack:1996vg,Jack:1996cn,Jack:1996qq,Harlander:2006xq}) were made in the $\overline{\mbox{DR}}$ scheme, when a theory is regularized by dimensional reduction \cite{Siegel:1979wq} and divergences are removed by modified minimal subtraction \cite{Bardeen:1978yd}. However, the dimensional reduction is not mathematically consistent \cite{Siegel:1980qs} and, as a result, can break supersymmetry in very higher loops \cite{Avdeev:1981vf,Avdeev:1982np,Avdeev:1982xy}. A much better regularization in the supersymmetric case is the higher covariant derivative method \cite{Slavnov:1971aw,Slavnov:1972sq,Slavnov:1977zf}. In the superfield formulation \cite{Krivoshchekov:1978xg,West:1985jx} (see also \cite{Aleshin:2016yvj,Kazantsev:2017fdc}) it is mathematically consistent, does not break supersymmetry in all orders, and, most importantly, allows revealing some special features of quantum corrections, see, e.g., \cite{Stepanyantz:2019lyo,Stepanyantz:2021qkk}. For instance, it was demonstrated that in the case of using the higher covariant derivative regularization the loop integrals giving the $\beta$-function are integrals of double total derivatives with respect to the loop momenta in all orders \cite{Stepanyantz:2011jy,Stepanyantz:2019ihw}.\footnote{The factorizations into total and double total derivatives were first noted in \cite{Soloshenko:2003nc} and \cite{Smilga:2004zr}, respectively. Subsequently, a similar structure of loop integrals was revealed in various non-Abelian supersymmetric theories regularized by higher covariant derivatives, see, e.g., \cite{Pimenov:2009hv,Stepanyantz:2011bz,Stepanyantz:2011cpt,Buchbinder:2014wra,Buchbinder:2015eva,Shakhmanov:2017soc,Kazantsev:2018nbl}. For theories regularized by dimensional reduction the factorization into integrals of (double) total derivatives does not take place \cite{Aleshin:2015qqc,Aleshin:2016rrr}.} This fact turns out to be the key to explaining the appearance of the NSVZ equation \cite{Novikov:1983uc,Jones:1983ip,Novikov:1985rd,Shifman:1986zi}, which relates the $\beta$-function of ${\cal N}=1$ supersymmetric gauge theories to the anomalous dimension of the matter superfields. Its perturbative proof in the Abelian case has been done in \cite{Stepanyantz:2011jy} and \cite{Stepanyantz:2014ima} by two different methods. A much more complicated proof in the non-Abelian case was constructed in \cite{Stepanyantz:2020uke}, see also \cite{Stepanyantz:2019lfm} on the base of the non-renormalization theorem for the triple gauge ghost vertices \cite{Stepanyantz:2016gtk}.\footnote{The higher covariant derivative regularization was essentially used for deriving this non-renormalization theorem in \cite{Stepanyantz:2016gtk} (see also \cite{Korneev:2021zdz} for the generalization to the case of theories with multiple gauge couplings) and making its perturbative check in the two-loop approximation \cite{Kuzmichev:2021yjo,Kuzmichev:2021lqa}.} It turned out that the NSVZ equation is valid in all loops for the renormalization group functions (RGFs) defined in terms of the bare couplings in the case of using the higher covariant derivative regularization for an arbitrary renormalization prescription. For standard RGFs, which are defined in terms of the renormalized couplings, the NSVZ equation is valid in the HD+MSL scheme \cite{Kataev:2013eta} when the theory is regularized by higher covariant derivatives and the renormalization is made with the help of minimal subtractions of logarithms \cite{Shakhmanov:2017wji,Stepanyantz:2017sqg}. In the Abelian case another all-loop NSVZ renormalization prescription is the on-shell scheme \cite{Kataev:2019olb}.

The higher covariant derivative regularization also allows deriving some NSVZ-like relations and constructing the corresponding all-loop NSVZ renormalization prescriptions. For example, the NSVZ-like relation for the Adler $D$-function in ${\cal N}=1$ SQCD \cite{Shifman:2014cya,Shifman:2015doa} (see also \cite{Kataev:2017qvk}) can also be derived using the method proposed in \cite{Stepanyantz:2011jy}. In the softly broken SQED the NSVZ-like relation for the renormalization of the gaugino mass \cite{Hisano:1997ua,Jack:1997pa,Avdeev:1997vx} can also be obtained by this method \cite{Nartsev:2016nym,Nartsev:2016mvn}.

Nevertheless, although the higher covariant derivative regularization has a lot of attractive features, its application is technically difficult due to the complicated structure of vertices coming from the higher derivative terms and complicated loop integrals. At present it was used only for calculating the two-loop anomalous dimension and the three-loop $\beta$-function, see, e.g., \cite{Aleshin:2020gec,Kazantsev:2020kfl}. Even in this case the calculation of loop integrals was rather non-trivial, see, e.g., \cite{Soloshenko:2002np,Soloshenko:2003nc}. Therefore, the question arises as to whether it is possible to use the higher covariant derivative regularization for making explicit calculations in higher orders.

In this paper we consider the ${\cal N}=1$ supersymmetric electrodynamics (SQED) with $N_f$ flavors (which is the simplest ${\cal N}=1$ supersymmetric gauge theory) and calculate for it the three-loop anomalous dimension of the matter superfields and the four-loop $\beta$-function. In Sect. \ref{Section_SQED_With_Nf_Flavors} we briefly describe how this theory is formulated and regularized by higher derivatives. After this in Sect. \ref{Section_Bare_RGFs} we obtain the anomalous dimension defined in terms of the bare coupling constant. The corresponding supergraphs were generated and reduced to the momentum integrals with the help of a special C++ computer programm written by I.S. The resulting integrals are calculated with the help of the Chebyshev polynomial technique \cite{Rosner:1967zz} in Appendix \ref{Appendix_Integrals_For_Gamma}. After obtaining the result for the three-loop anomalous dimension defined in terms of the bare coupling constant we construct the expression for the four-loop $\beta$-function defined in terms of the bare coupling constant. For this purpose we use the general statement  that RGFs defined in terms of the bare coupling constant satisfy the NSVZ equation in the case of using the higher covariant derivative regularization, see, e.g., \cite{Stepanyantz:2011jy}. Note that this statement is valid for an arbitrary renormalization prescription because RGFs defined in terms of the bare couplings are scheme independent for a fixed regularization, but, certainly, depend on a regularization. RGFs standardly defined in terms of the renormalized coupling constant are obtained in Sect. \ref{Section_Renormalized_RGFs}. They depend on a regularization and a renormalization prescription and satisfy the NSVZ equation only in some special subtractions schemes (which in particular include the HD+MSL and on-shell schemes). Analysing the scheme dependence we see that all terms proportional to $N_f$ in the three-loop anomalous dimension and proportional to $(N_f)^2$ in the four-loop $\beta$-function can be set to 0 by a special choice of a renormalization prescription. In the resulting subtraction scheme (which we will call ``the minimal scheme'') the expressions for RGFs have the simplest form and satisfy the NSVZ equation. In Sect. \ref{Section_Minimal_Scheme} we demonstrate that the minimal scheme exists in all orders. Note that the proof is also valid in the non-supersymmetric case, say, for QED with $N_f$ flavors or for a scalar QED with $N_f$ flavors. For ${\cal N}=1$ SQED we also prove that the minimal scheme is NSVZ in all orders.

\section{${\cal N}=1$ SQED with $N_f$ flavors}
\hspace*{\parindent}\label{Section_SQED_With_Nf_Flavors}

In ${\cal N}=1$ superspace the action of massless ${\cal N}=1$ SQED with $N_f$ flavors is written as

\begin{equation}\label{Action_SQED}
S = \frac{1}{4e_0^2} \mbox{Re} \int d^4x\, d^2\theta\, W^a W_a + \frac{1}{4} \int d^4x\, d^4\theta\, \sum\limits_{\alpha=1}^{N_f} \Big(\phi_\alpha^* e^{2V} \phi_\alpha + \widetilde\phi_\alpha^* e^{-2V} \widetilde\phi_\alpha\Big),
\end{equation}

\noindent
where $\phi_\alpha$ and $\widetilde\phi_\alpha$ are $N_f$ pairs of chiral matter superfields with opposite $U(1)$ charges, and $V$ is a real gauge superfield. In our notation the bare and renormalized coupling constants are denoted by $e_0$ and $e$, respectively.

We regularize the theory (\ref{Action_SQED}) by the higher covariant derivatives following the ideas proposed in \cite{Slavnov:1971aw,Slavnov:1972sq,Slavnov:1977zf} and generalized to the supersymmetric case. Here we mainly use the same conventions as in \cite{Aleshin:2020gec}, but, for completeness, briefly describe the details of the regularization and quantization. The regularization is introduced by adding the higher derivative terms to the action (\ref{Action_SQED}), after which the regularized action

\begin{equation}
S_{\mbox{\scriptsize reg}} = \frac{1}{4e_0^2} \mbox{Re} \int d^4x\, d^2\theta\, W^a R(\partial^2/\Lambda^2) W_a + \frac{1}{4} \int d^4x\, d^4\theta\, \sum\limits_{\alpha=1}^{N_f} \Big(\phi_\alpha^* e^{2V} \phi_\alpha + \widetilde\phi_\alpha^* e^{-2V} \widetilde\phi_\alpha\Big)
\end{equation}

\noindent
will include the regulator function $R(x)$. This function grows at infinity and is equal to 1 at $x=0$. Another similar regulator function appears in the gauge fixing term

\begin{equation}
S_{\mbox{\scriptsize gf}} = -\frac{1}{32\xi_0 e_0^2} \int d^4x\, d^4\theta\,D^2 V K(\partial^2/\Lambda^2) \bar D^2 V,
\end{equation}

\noindent
where $\xi_0$ is the bare gauge parameter. The minimal (Feynman) gauge corresponds to $\xi_0=1$ and $R(x) = K(x)$. However, we will make calculations for an arbitrary $\xi_0$ and $K(x) \ne R(x)$. The generating functional of the regularized theory

\begin{equation}
Z[\mbox{sources}]= \int DV\,\Big(\prod\limits_{\alpha=1}^{N_f} D\phi_\alpha D\widetilde\phi_\alpha\Big)\, \mbox{Det}(PV, M)^{N_f} \exp\Big(iS_{\mbox{\scriptsize reg}} + iS_{\mbox{\scriptsize gf}} + iS_{\mbox{\scriptsize sources}}\Big)
\end{equation}

\noindent
also contains the Pauli--Villars determinant needed for removing residual one-loop divergences,

\begin{equation}
\mbox{Det}(PV,M)^{-1} = \int D\Phi\, D\widetilde\Phi\, \exp(iS_\Phi).
\end{equation}

\noindent
Here the action for the massive Pauli--Villars superfields is given be the expression

\begin{equation}
S_\Phi = \frac{1}{4}\int d^4x\,d^4\theta\, \Big(\Phi^* e^{2V} \Phi + \widetilde\Phi^* e^{-2V} \widetilde\Phi\Big) + \Big(\frac{M}{2}\int d^4x\, d^2\theta\, \widetilde\Phi\, \Phi  +\mbox{c.c.}\Big),
\end{equation}

\noindent
and it is important that the ratio of the Pauli--Villars mass $M$ to the regularization parameter $\Lambda$ should not depend on the coupling constant,

\begin{equation}\label{A_Small_Definition}
a \equiv \frac{M}{\Lambda} \ne a(e_0).
\end{equation}

Due to the renormalizability of the considered theory the ultraviolet divergences can be absorbed into the renormalization of the coupling constant, of the gauge parameter, and of the chiral matter superfields $\phi_\alpha$ and $\widetilde\phi_\alpha$. Note that all chiral superfields have the same renormalization constant $Z$, such that $\phi_{\alpha,R} = \sqrt{Z} \phi_\alpha$, $\widetilde\phi_{\alpha,R} = \sqrt{Z}\widetilde\phi_\alpha$ for all values of $\alpha=1,\ldots, N_f$.

It is convenient to encode the ultraviolet divergences in RGFs. Following \cite{Kataev:2013eta}, we distinguish between RGFs defined in terms of the bare coupling constant,

\begin{equation}\label{RGFs_Bare}
\beta(\alpha_0) = \frac{d\alpha_0}{d\ln\Lambda}\bigg|_{\alpha=\mbox{\scriptsize const}}; \qquad \gamma(\alpha_0) = - \frac{d\ln Z}{d\ln\Lambda}\bigg|_{\alpha=\mbox{\scriptsize const}},
\end{equation}

\noindent
and the ones standardly defined in terms of the renormalized coupling constant by the equations

\begin{equation}\label{RGFs_Renormalized}
\widetilde\beta(\alpha) = \frac{d\alpha}{d\ln\mu}\bigg|_{\alpha_0=\mbox{\scriptsize const}}; \qquad \widetilde\gamma(\alpha) = \frac{d\ln Z}{d\ln\mu}\bigg|_{\alpha_0=\mbox{\scriptsize const}},
\end{equation}

\noindent
where $\mu$ is a renormalization point.

RGFs (\ref{RGFs_Bare}) are independent of a renormalization prescription for a fixed regularization, but depend on a regularization. According to \cite{Stepanyantz:2011jy,Stepanyantz:2014ima} in the case of using the higher derivative regularization described above they satisfy the NSVZ equation \cite{Vainshtein:1986ja,Shifman:1985fi}

\begin{equation}\label{NSVZ_For_Bare_RGFs}
\frac{\beta(\alpha_0)}{\alpha_0^2} = \frac{N_f}{\pi}\Big(1-\gamma(\alpha_0)\Big)
\end{equation}

\noindent
in all loops for an arbitrary renormalization prescription. (In fact, both parts of this equation do not depend on this prescription in accordance with what was mentioned above.)

In contrast, RGFs defined by Eq. (\ref{RGFs_Renormalized}) depend on both a regularization and a subtraction scheme. For them the equation similar to (\ref{NSVZ_For_Bare_RGFs}) is valid only for some certain renormalization prescriptions called ``the NSVZ schemes''. Some of them are given by the HD+MSL renormalization prescription \cite{Kataev:2013eta,Kataev:2013csa,Kataev:2014gxa}, when the theory is regularized by higher derivatives and divergences are removed by minimal subtractions of logarithms \cite{Shakhmanov:2017wji,Stepanyantz:2017sqg}.

\section{The three-loop anomalous dimension and the four-loop $\beta$-function defined in terms of the bare coupling constant}
\hspace*{\parindent}\label{Section_Bare_RGFs}

Calculating superdiagrams with two external lines of the matter superfields one can obtain the function $G$ related to the corresponding part of the effective action by the equation

\begin{equation}
\Gamma_\phi^{(2)} = \frac{1}{4} \int \frac{d^4p}{(2\pi)^4}\, d^4\theta\,  \sum\limits_{\alpha=1}^{N_f}\Big(\phi_\alpha^*(p,\theta)\phi_\alpha(-p,\theta) + \widetilde\phi_\alpha^*(p,\theta) \widetilde\phi_\alpha(-p,\theta)\Big) G(\alpha_0,\Lambda/p).
\end{equation}

\noindent
If the function $G$ is known, then the anomalous dimension defined in terms of the bare coupling constant can be obtained with the help of the equation

\begin{equation}
\gamma(\alpha_0) = \frac{d\ln G}{d\ln\Lambda}\bigg|_{\alpha=\mbox{\scriptsize const};\, p\to 0},
\end{equation}

\noindent
where the condition $p\to 0$ removes terms proportional to powers of $p/\Lambda$.

The superdiagrams contributing to the three-loop anomalous dimension of the matter superfields were generated by a special C++ program written by I.S. Also this program calculated the $D$-algebra and reduced the expression for a superdiagram to a momentum integral using the standard technique for calculating supergraphs, see, e.g., \cite{Gates:1983nr,West:1990tg,Buchbinder:1998qv}. The result obtained with the help of this program is written as

\begin{eqnarray}\label{Gamma_Original_Integrals}
&&\hspace*{-5mm} \gamma(\alpha_0) = \frac{d\ln G}{d\ln\Lambda}\bigg|_{P=0} = - \frac{d}{d\ln\Lambda} \int \frac{d^4K}{(2\pi)^4} \frac{2e_0^2}{K^4 R_K}
+ \frac{d}{d\ln\Lambda} \int \frac{d^4K}{(2\pi)^4} \frac{d^4L}{(2\pi)^4} \frac{e_0^4}{R_K R_L} \bigg(\frac{4}{K^2 L^4 (K+L)^2}\nonumber\\
&&\hspace*{-5mm} - \frac{2}{K^4 L^4} \bigg)
+ N_f \frac{d}{d\ln\Lambda} \int \frac{d^4K}{(2\pi)^4} \frac{d^4L}{(2\pi)^4} \frac{4e_0^4}{R_K^2 K^4} \bigg(\frac{1}{L^2 (L+K)^2} - \frac{1}{(L^2+M^2)((L+K)^2+M^2)}\bigg)\nonumber\\
&&\hspace*{-5mm} + \frac{d}{d\ln\Lambda} \int \frac{d^4K}{(2\pi)^4} \frac{d^4L}{(2\pi)^4} \frac{d^4Q}{(2\pi)^4} \frac{8 e_0^6}{R_K R_L R_Q}\bigg[- \frac{1}{3K^4 L^4 Q^4} + \frac{1}{K^4 L^2 Q^4 (Q+L)^2} + \frac{1}{K^2 L^4 (K+L)^2}\nonumber\\
&&\hspace*{-5mm} \times \frac{1}{(Q+K+L)^2}\bigg(\frac{1}{Q^2} - \frac{2}{(Q+L)^2}\bigg) \bigg]
+ N_f \frac{d}{d\ln\Lambda} \int \frac{d^4K}{(2\pi)^4} \frac{d^4L}{(2\pi)^4} \frac{d^4Q}{(2\pi)^4} \frac{16 e_0^6\, K_\mu L^\mu}{R_K^2 R_L K^4L^4 (K+L)^2}\nonumber\\
&&\hspace*{-5mm} \times \bigg(\frac{1}{Q^2 (Q+K)^2} - \frac{1}{(Q^2+M^2)((Q+K)^2+M^2)}\bigg)
+ N_f \frac{d}{d\ln\Lambda} \int \frac{d^4K}{(2\pi)^4} \frac{d^4L}{(2\pi)^4} \frac{d^4Q}{(2\pi)^4} \frac{8e_0^6}{R_K^2 R_L K^4}\nonumber\\
&&\hspace*{-5mm} \times\frac{1}{L^2}\bigg(\frac{2(Q+K+L)^2-K^2-L^2}{Q^2(Q+K)^2(Q+L)^2(Q+K+L)^2}
- \frac{2(Q+K+L)^2 - K^2 - L^2}{(Q^2+M^2)((Q+K)^2+M^2)((Q+L)^2+M^2)}\nonumber\\
&&\hspace*{-5mm} \times \frac{1}{((Q+K+L)^2+M^2)} + \frac{4M^2}{(Q^2+M^2)^2 ((Q+K)^2+M^2) ((Q+L)^2+M^2)}
\bigg)
- (N_f)^2 \frac{d}{d\ln\Lambda} \nonumber\\
&&\hspace*{-5mm} \times \int \frac{d^4K}{(2\pi)^4} \frac{d^4L}{(2\pi)^4} \frac{d^4Q}{(2\pi)^4} \frac{8 e_0^6}{R_K^3 K^4}\bigg(\frac{1}{Q^2 (Q+K)^2} - \frac{1}{(Q^2+M^2)((Q+K)^2 +M^2)}\bigg) \bigg(\frac{1}{L^2 (L+K)^2}\nonumber\\
&&\hspace*{-5mm} - \frac{1}{(L^2+M^2)((L+K)^2 +M^2)}\bigg) + O(e_0^8),
\end{eqnarray}

\noindent
where $R_K \equiv R(K^2/\Lambda^2)$. Note that the calculation was made for an arbitrary value of the gauge parameter $\xi_0$, but the gauge dependence disappeared as should be according to the general theorems, see \cite{Batalin:2019wkb} and references therein. Certainly, this fact can be considered as a useful correctness check.

Calculating the expression (\ref{Gamma_Original_Integrals}) one should express the bare coupling constant in terms of the renormalized one and only after this make the differentiation with respect to $\ln\Lambda$, which should certainly be done before integrations. In the lowest orders the bare and renormalized coupling constants ($\alpha_0 = e_0^2/4\pi$ and $\alpha= e^2/4\pi$, respectively) are related by the equation (see, e.g., \cite{Aleshin:2020gec})

\begin{equation}\label{Relation_Between_Couplings}
\frac{1}{\alpha_0} = \frac{1}{\alpha} - \frac{N_f}{\pi} \Big(\ln\frac{\Lambda}{\mu} + b_{1,0}\Big) - \frac{\alpha N_f}{\pi^2}\Big(\ln\frac{\Lambda}{\mu} + b_{2,0}\Big) + O(\alpha^2),
\end{equation}

\noindent
where $b_{1,0}$ and $b_{2,0}$ are finite constants which (partially) determine a subtraction scheme in the one- and two-loop approximations, respectively. Note that similar finite constants appear in the renormalization constant for the chiral matter superfields. For instance, in the lowest approximation

\begin{equation}
\ln Z = \frac{\alpha}{\pi}\Big(\ln\frac{\Lambda}{\mu} + g_{1,0}\Big) + O(\alpha^2).
\end{equation}

The integrals present in the expression (\ref{Gamma_Original_Integrals}) (rewritten in terms of the renormalized coupling constant with the help of Eq. (\ref{Relation_Between_Couplings})) are calculated in Appendix \ref{Appendix_Integrals_For_Gamma} using the Chebyshev polynomials method \cite{Rosner:1967zz}. The Chebyshev polynomials are defined as

\begin{equation}
C_n(\cos\theta) \equiv \frac{\sin\, ((n+1)\theta)}{\sin\theta}
\end{equation}

\noindent
and (for $t<1$) satisfy the important equation

\begin{equation}
\frac{1}{1-2tz + t^2} = \sum\limits_{n=0}^\infty t^n C_n(z).
\end{equation}

\noindent
Consequently, the function $(K-L)^{-2} = (K^2 - 2 KL\cos\theta+L^2)^{-1}$, where $\theta$ is the angle between the Euclidian four-vectors $K_\mu$ and $L_\mu$, can be presented in the form

\begin{equation}\label{Inverse_Square}
\frac{1}{(K-L)^2} = \left\{\begin{array}{l}
{\displaystyle \frac{1}{K^2} \sum\limits_{n=0}^\infty \Big(\frac{L}{K}\Big)^n C_n(\cos\theta),\quad\mbox{if}\quad K > L;}\\
\vphantom{1}\\
{\displaystyle \frac{1}{L^2} \sum\limits_{n=0}^\infty \Big(\frac{K}{L}\Big)^n C_n(\cos\theta),\quad\, \mbox{if}\quad L > K.}
\end{array}
\right.
\end{equation}

\noindent
Then the angular parts of the loops integrals can be calculated with the help of the useful identities

\begin{eqnarray}\label{Product}
&& \int \frac{d\Omega_Q}{2\pi^2} C_m\Big(\frac{K_\mu Q^\mu}{KQ}\Big) C_n\Big(\frac{Q_\nu L^\nu}{QL}\Big) = \frac{1}{n+1} \delta_{mn} C_n\Big(\frac{K_\mu L^\mu}{KL}\Big);\\
\label{Ortogonality}
&& \int \frac{d\Omega}{2\pi^2} C_m(\cos \theta) C_n(\cos \theta) = \delta_{mn},
\end{eqnarray}

\noindent
where $d\Omega$ is the element of a solid angle on a sphere $S^3$ in the momentum space. (The subscript $Q$ indicates that this sphere lives in the space with the Cartesian coordinates $Q_1,Q_2,Q_3,Q_4$.)

The calculation of the integrals in Eq. (\ref{Gamma_Original_Integrals}) is described in Appendix \ref{Appendix_Integrals_For_Gamma}. The resulting expression for the three-loop anomalous dimension of the matter superfields defined in terms of the bare coupling constant is written as

\begin{eqnarray}\label{Gamma_3Loop_Bare}
&& \gamma(\alpha_0) = - \frac{\alpha_0}{\pi} + \frac{\alpha_0^2}{2\pi^2} + \frac{\alpha_0^2 N_f}{\pi^2}\Big(\ln a + 1 + \frac{A_1}{2}\Big) - \frac{\alpha_0^3}{2\pi^3} - \frac{\alpha_0^3 N_f}{\pi^3}\Big(\ln a +\frac{3}{4} +C\Big)\nonumber\\
&& - \frac{\alpha_0^3(N_f)^2}{\pi^3} \Big((\ln a + 1)^2 -\frac{A_2}{4} + D_1 \ln a + D_2\Big) + O(\alpha_0^4),
\end{eqnarray}

\noindent
where $A_1$, $A_2$, $C$, $D_1$, and $D_2$ are numerical parameters depending on the regulator function $R(x)$. They are defined by the following equations:

\begin{eqnarray}\label{Constants_Definition}
&& \hspace*{-7mm} A_1 \equiv \int\limits_0^\infty dx\, \ln x\, \frac{d}{dx}\Big(\frac{1}{R(x)}\Big);\qquad\qquad\qquad\qquad\qquad
A_2 \equiv \int\limits_0^\infty dx\, \ln^2 x\, \frac{d}{dx}\Big(\frac{1}{R(x)}\Big);\nonumber\\
&& \hspace*{-7mm} C \equiv \int\limits_0^1 dx \int\limits_0^\infty dy\, x\, \ln y\, \frac{d}{dy}\Big(\frac{1}{R(y) R(x^2 y)}\Big);\qquad\qquad\ D_1 \equiv \int\limits_0^\infty dx\, \ln x\, \frac{d}{dx}\Big(\frac{1}{R^2(x)}\Big);\nonumber\\
&& \hspace*{-7mm} D_2 \equiv \int\limits_0^\infty dx\, \ln x\, \frac{d}{dx}\left\{\frac{1}{R^2(x)}\bigg[-\frac{1}{2}\Big(1-R(x)\Big) \ln x + \sqrt{1+\frac{4a^2}{x}}\, \mbox{arctanh}\sqrt{\frac{x}{x+4a^2}}\,\bigg]\right\}.
\end{eqnarray}

\noindent
Note that the expression (\ref{Gamma_3Loop_Bare}) depends only on the regularization parameters, but is independent of the renormalization parameters which fix a subtraction scheme in the considered approximation. Certainly, this occurs because we consider the anomalous dimension defined in terms of the bare coupling constant, which is scheme independent for a fixed regularization \cite{Kataev:2013eta}.

In the particular case $R(x) = 1+ x^n$ some of the regularization parameters can be calculated analytically,

\begin{eqnarray}\label{ABCD1_Values}
&& A_1\Big|_{R(x) = 1+ x^n} = 0;\qquad\qquad\qquad\qquad\ \, A_2\Big|_{R(x) = 1+ x^n} = - \frac{\pi^2}{3n^2};\qquad \nonumber\\
&& C\Big|_{R(x) = 1+ x^n} = \frac{1}{2n^2}\psi'\Big(\frac{n+1}{n}\Big);\qquad\quad D_1\Big|_{R(x) = 1+ x^n} = \frac{1}{n},
\end{eqnarray}

\noindent
where $\psi(z)$ is the logarithmic derivative of the gamma function $\Gamma(z)$,

\begin{equation}
\psi(z) \equiv \frac{d\ln\Gamma(z)}{dz}.
\end{equation}

\noindent
For $n=1,2$ the expression for the constant $C$ can be simplified,

\begin{equation}\label{C_Values}
C\Big|_{R(x) = 1+ x} = \frac{\pi^2}{12}-\frac{1}{2}; \qquad  C\Big|_{R(x) = 1+ x^2} = \frac{\pi^2}{16}-\frac{1}{2}.
\end{equation}

\noindent
Unfortunately we did manage to obtain an analytic result for the constant $D_2$ even for $R(x) = 1+x^n$, although in this case

\begin{equation}
\int\limits_0^\infty dx\, \ln x\, \frac{d}{dx}\bigg(\frac{ x^n \ln x}{2(1+x^n)^2}\bigg) = 0.
\end{equation}

\noindent
However, this constant can easily be calculated numerically, for instance,

\begin{equation}\label{D2_Values}
D_2\Big|_{R(x) = 1+ x;\ a=1} \approx 0.93; \qquad  D_2\Big|_{R(x) = 1+ x^2;\ a=1} \approx 0.44.
\end{equation}

To find the $\beta$-function defined in terms of the bare coupling constant, we substitute the expression (\ref{Gamma_3Loop_Bare}) into the NSVZ equation (\ref{NSVZ_For_Bare_RGFs}). According to \cite{Stepanyantz:2011jy,Stepanyantz:2014ima}, for RGFs defined in terms of the bare coupling constant it is valid in all loops for an arbitrary renormalization prescription supplementing the higher derivative regularization.  Therefore, the four-loop $\beta$-function takes the form

\begin{eqnarray}
&& \beta(\alpha_0) = \frac{\alpha_0^2 N_f}{\pi} + \frac{\alpha_0^3 N_f}{\pi^2} - \frac{\alpha_0^4 N_f}{2\pi^3} - \frac{\alpha_0^4 (N_f)^2}{\pi^3}\Big(\ln a + 1 + \frac{A_1}{2}\Big) + \frac{\alpha_0^5 N_f}{2\pi^4}
+ \frac{\alpha_0^5 (N_f)^2}{\pi^4} \qquad\nonumber\\
&& \times \Big(\ln a + \frac{3}{4} + C\Big) + \frac{\alpha_0^5(N_f)^3}{\pi^4} \Big((\ln a + 1)^2 -\frac{A_2}{4} + D_1 \ln a + D_2\Big) + O(\alpha_0^6).\qquad
\end{eqnarray}

\noindent
Certainly, as any renormalization group function defined in terms of the bare couplings, this expression depends only on the regularization parameters, but is independent of the parameters which determine a subtraction scheme (for a fixed regularization).

\section{RGFs defined in terms of the renormalized coupling constant}
\label{Section_Renormalized_RGFs}

\subsection{The algorithm for obtaining RGFs defined in terms of the renormalized coupling constant}
\hspace*{\parindent}\label{Subsection_Algorithm}

Starting from RGFs defined in terms of the bare coupling constant one can obtain RGFs defined in terms of the renormalized coupling constant. For this purpose it is necessary to do the following:

1. Using the expressions for $\alpha(\alpha_0)$ and $\ln Z$ in the previous orders, the functions $\beta(\alpha_0)$ and $\gamma(\alpha_0)$ should be rewritten in terms of the renormalized coupling constant $\alpha$. The results are substituted into the left hand side of the equations (\ref{RGFs_Bare}).

2. Integrating the equations (\ref{RGFs_Bare}) we obtain the functions $\alpha_0(\alpha,\ln\Lambda/\mu)$ and $\ln Z(\alpha,\ln\Lambda/\mu)$. They will contain some integration constants which fix a subtraction scheme in the considered approximation.

3. Next, it is necessary to construct the functions $\alpha(\alpha_0,\Lambda/\mu)$ and $\ln Z(\alpha_0,\ln\Lambda/\mu)\equiv \ln Z(\alpha(\alpha_0,\ln\Lambda/\mu),\ln\Lambda/\mu)$.

4. The functions $\alpha(\alpha_0,\Lambda/\mu)$ and $\ln Z(\alpha_0,\Lambda/\mu)$ are substituted into the equations (\ref{RGFs_Renormalized}).

5. Resulting RGFs should be written in terms of the renormalized coupling constant. After this all $\ln\Lambda/\mu$ should cancel each other providing a test of the calculation correctness.

\subsection{The three-loop anomalous dimension defined in terms of the renormalized coupling constant}
\hspace*{\parindent}\label{Subsection_Gamma_Renormalized}

For the anomalous dimension of the matter superfields this procedure is as follows: Integrating the equation

\begin{equation}
\frac{d\ln Z}{d\ln\Lambda}\bigg|_{\alpha=\mbox{\scriptsize const}} = - \gamma(\alpha_0)
\end{equation}

\noindent
we obtain the expression for $\ln Z$ written in terms of the renormalized coupling constant,

\begin{eqnarray}\label{Ln_Z}
&&\hspace*{-5mm} \ln Z = \frac{\alpha}{\pi}\Big(\ln\frac{\Lambda}{\mu} + g_{1,0}\Big) - \frac{\alpha^2}{2\pi^2} \Big(\ln \frac{\Lambda}{\mu} + g_{2,0} + N_f g_{2,1}\Big) - \frac{\alpha^2 N_f}{\pi^2} \Big(\ln a + 1 + \frac{A_1}{2} - b_{1,0}\Big) \ln \frac{\Lambda}{\mu}\nonumber\\
&&\hspace*{-5mm} + \frac{\alpha^2 N_f}{2\pi^2} \ln^2\frac{\Lambda}{\mu} + \frac{\alpha^3}{2\pi^3}\Big(\ln\frac{\Lambda}{\mu} + g_{3,0} + N_f g_{3,1} + (N_f)^2 g_{3,2}\Big)
+ \frac{\alpha^3 N_f}{\pi^3} \Big(\ln a + \frac{3}{4} + C + b_{2,0} - b_{1,0} \Big) \nonumber\\
&&\hspace*{-5mm} \times \ln\frac{\Lambda}{\mu} + \frac{\alpha^3 (N_f)^2}{\pi^3} \Big(\frac{1}{3}\ln^3\frac{\Lambda}{\mu} + b_{1,0} \ln^2\frac{\Lambda}{\mu} + (b_{1,0})^2 \ln\frac{\Lambda}{\mu}\Big) - \frac{\alpha^3 (N_f)^2}{\pi^3} \Big(\ln a +1 +\frac{A_1}{2}\Big)\Big(\ln^2\frac{\Lambda}{\mu} \nonumber\\
&&\hspace*{-5mm} + 2 b_{1,0} \ln\frac{\Lambda}{\mu}\Big) + \frac{\alpha^3 (N_f)^2}{\pi^3} \Big((\ln a + 1)^2 - \frac{A_2}{4} + D_1 \ln a + D_2\Big) \ln \frac{\Lambda}{\mu} + O(\alpha^4).
\end{eqnarray}

\noindent
Note that in this expression the finite constants fixing a subtraction scheme are introduced in such a way that the terms containing them will be polynomials in $N_f$ with the same maximal power as the quantum correction in the corresponding approximation. This implies that the renormalization prescription is compatible with the structure of quantum corrections. Rewriting $\ln Z$ in terms of the bare coupling constant and differentiating the result with respect to $\ln\mu$ we construct the anomalous dimension defined in terms of the renormalized coupling constant,

\begin{eqnarray}\label{Gamma3_Result}
&&\hspace*{-5mm} \widetilde\gamma(\alpha)= - \frac{\alpha}{\pi} + \frac{\alpha^2}{2\pi^2} + \frac{\alpha^2 N_f}{\pi^2}\Big(\ln a + 1 + \frac{A_1}{2} + g_{1,0} - b_{1,0}\Big) - \frac{\alpha^3}{2\pi^3}
+ \frac{\alpha^3 N_f}{\pi^3} \Big(- \ln a - \frac{3}{4} - C   \nonumber\\
&&\hspace*{-5mm} - b_{2,0} + b_{1,0} - g_{2,0} + g_{1,0}\Big) + \frac{\alpha^3 (N_f)^2}{\pi^3}\Big\{-\Big(\ln a + 1 - b_{1,0}\Big)^2 + \frac{A_2}{4} - D_1 \ln a - D_2 + b_{1,0} A_1\nonumber\\
&&\hspace*{-5mm} - g_{2,1}\Big\} + O(\alpha^4). \vphantom{\frac{N_f^2}{\pi^2}}
\end{eqnarray}

\noindent
For the values of finite constants

\begin{eqnarray}
&& g_{1,0} = - \ln a -\frac{1}{2} - \frac{A_1}{2} + b_{1,0};\nonumber\\
&& g_{2,0} = -\ln a - C + g_{1,0} + b_{1,0} - b_{2,0}-\frac{7}{4} +\frac{3}{2}\zeta(3);\nonumber\\
&& g_{2,1} = -\Big(\ln a + 1 - b_{1,0}\Big)^2 + \frac{1}{4} (A_2-1) - D_1 \ln a - D_2 + b_{1,0} A_1 \qquad
\end{eqnarray}

\noindent
the expression (\ref{Gamma3_Result}) produces the result in the $\overline{\mbox{DR}}$ scheme,

\begin{eqnarray}\label{Gamma3_DR}
&&\hspace*{-5mm} \widetilde\gamma_{\overline{\mbox{\scriptsize DR}}}(\alpha)= - \frac{\alpha}{\pi} + \frac{\alpha^2}{2\pi^2} + \frac{\alpha^2 N_f}{2\pi^2} - \frac{\alpha^3}{2\pi^3}
+ \frac{\alpha^3 N_f}{\pi^3} \Big(1-\frac{3}{2}\zeta(3)\Big) + \frac{\alpha^3 (N_f)^2}{4\pi^3} + O(\alpha^4),
\end{eqnarray}

\noindent
which was found in Ref. \cite{Jack:1996qq}. Note that the terms which do not contain $N_f$ are scheme independent and should be the same in all subtraction schemes \cite{Kataev:2013csa}. The coincidence of these terms in Eqs. (\ref{Gamma3_Result}) and (\ref{Gamma3_DR}) can be considered as a test of the calculation correction.

Another interesting observation is that the finite constants which determine the renormalization of the matter superfields can be chosen so as to set all scheme dependent terms (which are proportional to $(N_f)^k$ with $k\ge 1$) to 0. Really, if

\begin{eqnarray}\label{Minimal_Scheme_Original_Constants_G}
&& g_{1,0} = - \ln a - 1 - \frac{A_1}{2} + b_{1,0};\nonumber\\
&& g_{2,0} = -\ln a - \frac{3}{4} - C + g_{1,0} + b_{1,0} - b_{2,0};\nonumber\\
&& g_{2,1} =  -\Big(\ln a + 1 - b_{1,0}\Big)^2 + \frac{A_2}{4} - D_1 \ln a - D_2 + b_{1,0} A_1, \qquad
\end{eqnarray}

\noindent
then all scheme dependent terms vanish and the anomalous dimension of the matter superfields takes the simplest possible form

\begin{equation}\label{Gamma_Minimal}
\widetilde\gamma(\alpha) = - \frac{\alpha}{\pi} + \frac{\alpha^2}{2\pi^2} - \frac{\alpha^3}{2\pi^3} + O(\alpha^4).
\end{equation}

\noindent
Below this renormalization prescription will be called ``the minimal scheme''. Actually, by a special tuning of a renormalization prescription we obtained the result coinciding with the one in the so-called ``conformal symmetry limit'' if we adopt the terminology of Ref. \cite{Kataev:2013vua}. However, instead of omitting certain supergraphs here we set the corresponding contributions to 0 by a proper choice of finite constants fixing a subtraction scheme.

Note that one can make such a finite renormalization that all contributions to the anomalous dimension beyond the one-loop approximation vanish. This choice corresponds to the so-called ``t'Hooft scheme'' \cite{tHooft:1977ysd,tHooft:1977xjm}, see also \cite{Garkusha:2011xb} and references therein. For example, the two- and three-loop contributions to the anomalous dimension can be set to 0 if we choose

\begin{eqnarray}
&& g_{1,0} = - \frac{1}{2N_f}-\ln a -1 -\frac{A_1}{2} + b_{1,0};\nonumber\\
&& g_{2,0} + N_f g_{2,1} = -\frac{1}{2N_f}  -\ln a -\frac{3}{4} - C + g_{1,0} + b_{1,0} - b_{2,0} \nonumber\\
&&\qquad\qquad\quad + N_f\Big\{-\Big(\ln a + 1 - b_{1,0}\Big)^2 + \frac{A_2}{4} - D_1 \ln a - D_2 + b_{1,0} A_1\Big\}.\qquad
\end{eqnarray}

\noindent
The presence of the terms proportional to $(N_f)^{-1}$ in these expressions tells us that the finite renormalization which gives the t'Hooft scheme is not compatible with the structure of quantum corrections, because such a dependence on $N_f$ cannot appear in any supergraph contributing to the considered renormalization group function.

Therefore, it is Eq. (\ref{Gamma_Minimal}) that is the simplest expression for the anomalous dimensions among those that can be obtained by finite renormalizations compatible with the structure of quantum corrections.

\subsection{The four-loop $\beta$-function defined in terms of the renormalized coupling constant}
\hspace*{\parindent}\label{Subsection_Beta_Renormalized}

To obtain the four-loop $\beta$-function defined in terms of the renormalized coupling constant, we first integrate the equation

\begin{equation}
\frac{d}{d\ln\Lambda}\Big(\frac{1}{\alpha_0}\Big)\bigg|_{\alpha=\mbox{\scriptsize const}} = - \frac{\beta(\alpha_0)}{\alpha_0^2}.
\end{equation}

\noindent
As a result we obtain the dependence $\alpha_0(\alpha)$ which again contains some finite constants which specify a subtraction scheme in the considered approximation,

\begin{eqnarray}\label{Full_Couplings_Relations}
&&\hspace*{-5mm} \frac{1}{\alpha_0}  = \frac{1}{\alpha} - \frac{N_f}{\pi} \Big(\ln\frac{\Lambda}{\mu} + b_{1,0}\Big) - \frac{\alpha N_f}{\pi^2}\Big(\ln\frac{\Lambda}{\mu} + b_{2,0}\Big)
+ \frac{\alpha^2 N_f}{2\pi^3}\Big( \ln\frac{\Lambda}{\mu} + b_{3,0} + N_f b_{3,1}\Big) \nonumber\\
&&\hspace*{-5mm} + \frac{\alpha^2 (N_f)^2}{\pi^3}\Big(\ln a + 1 + \frac{A_1}{2} - b_{1,0}\Big)\ln \frac{\Lambda}{\mu} - \frac{\alpha^2 (N_f)^2}{2\pi^3}\ln^2\frac{\Lambda}{\mu} -\frac{\alpha^3 N_f}{2\pi^4}\Big(\ln\frac{\Lambda}{\mu} + b_{4,0} + N_f b_{4,1} \nonumber\\
&&\hspace*{-5mm} + (N_f)^2 b_{4,2}\Big) - \frac{\alpha^3 (N_f)^2}{\pi^4} \Big(\ln a + \frac{3}{4} + C + b_{2,0} - b_{1,0}\Big)\ln\frac{\Lambda}{\mu} - \frac{\alpha^3 (N_f)^3}{\pi^4}\Big(\frac{1}{3}\ln^3\frac{\Lambda}{\mu} + b_{1,0} \nonumber\\
&&\hspace*{-5mm} \times \ln^2\frac{\Lambda}{\mu} + (b_{1,0})^2 \ln\frac{\Lambda}{\mu}\Big) + \frac{\alpha^3 (N_f)^3}{\pi^4} \Big(\ln a + 1 + \frac{A_1}{2}\Big)\Big(\ln^2\frac{\Lambda}{\mu} + 2 b_{1,0}\ln\frac{\Lambda}{\mu}\Big) - \frac{\alpha^3 (N_f)^3}{\pi^4}\nonumber\\
&&\hspace*{-5mm} \times \Big((\ln a +1)^2 - \frac{A_2}{4} + D_1 \ln a + D_2\Big) \ln\frac{\Lambda}{\mu} + O(\alpha^4). \vphantom{\frac{1}{2}}
\end{eqnarray}

\noindent
Solving this equation for $1/\alpha$ and differentiating the result with respect to $\ln\mu$ we construct the required $\beta$-function,

\begin{eqnarray}\label{Beta4_Result}
&&\hspace*{-5mm} \frac{\widetilde\beta(\alpha)}{\alpha^2} = -\frac{d}{d\ln\mu}\Big(\frac{1}{\alpha}\Big)\bigg|_{\alpha_0 = \mbox{\scriptsize const}} = \frac{N_f}{\pi} + \frac{\alpha N_f}{\pi^2} - \frac{\alpha^2 N_f}{2\pi^3} - \frac{\alpha^2 (N_f)^2}{\pi^3}\Big(\ln a + 1+\frac{A_1}{2} + b_{2,0} - b_{1,0}\Big)
\nonumber\\
&&\hspace*{-5mm} + \frac{\alpha^3 N_f}{2\pi^4} + \frac{\alpha^3 (N_f)^2}{\pi^4} \Big(\ln a + \frac{3}{4} + C + b_{3,0} - b_{1,0}\Big) + \frac{\alpha^3 (N_f)^3}{\pi^4} \Big\{ \Big(\ln a +1 - b_{1,0}\Big)^2 - b_{1,0} A_1
+ b_{3,1} \nonumber\\
&&\hspace*{-5mm}  -\frac{A_2}{4} + D_1 \ln a + D_2   \Big\} + O(\alpha^4).\vphantom{\frac{1}{2}}
\end{eqnarray}

\noindent
The $\overline{\mbox{DR}}$ result \cite{Jack:1996cn}

\begin{eqnarray}\label{Beta4_DR}
&& \frac{\widetilde\beta_{\overline{\mbox{\scriptsize DR}}}(\alpha)}{\alpha^2} = \frac{N_f}{\pi} + \frac{\alpha N_f}{\pi^2} - \frac{\alpha^2 N_f}{2\pi^3} - \frac{3\alpha^2 (N_f)^2}{4\pi^3}\nonumber\\
&&\qquad\qquad\qquad\quad + \frac{\alpha^3 N_f}{2\pi^4} + \frac{\alpha^3 (N_f)^2}{\pi^4}\Big(-\frac{5}{6}+\frac{3}{2}\zeta(3)\Big) + \frac{\alpha^3 (N_f)^3}{12\pi^4} + O(\alpha^4)\qquad
\end{eqnarray}

\noindent
is obtained from this expression if the finite constants which determine a subtraction scheme take the values

\begin{eqnarray}
&& b_{2,0} =  - \ln a - \frac{1}{4} - \frac{A_1}{2} + b_{1,0};\nonumber\\
&& b_{3,0} = - \ln a - C + b_{1,0} - \frac{19}{12}+\frac{3}{2}\zeta(3);\nonumber\\
&& b_{3,1} = -\Big(\ln a + 1 - b_{1,0}\Big)^2 + b_{1,0} A_1 + \frac{A_2}{4} - D_1 \ln a - D_2 + \frac{1}{12}.\qquad
\end{eqnarray}

\noindent
Certainly, all scheme independent terms in Eq. (\ref{Beta4_Result}) and (\ref{Beta4_DR}) coincide. However, this is rather evident, because such terms coincided in the expression for the three-loop anomalous dimension. Really, the scheme independent terms in the $L$-loop $\beta$-function and in the $(L-1)$-loop anomalous dimension are related by the NSVZ equation. In some schemes the NSVZ equation takes place according to the results of \cite{Stepanyantz:2011jy,Kataev:2013eta}. Therefore, it should be valid for the terms which do not depend on a renormalization presciption.

Also we see that all scheme dependent terms can be set to 0 if the finite constants are chosen so as to satisfy the equations

\begin{eqnarray}\label{Minimal_Scheme_Original_Constants_B}
&& b_{2,0} =  - \ln a - 1 - \frac{A_1}{2} + b_{1,0};\nonumber\\
&& b_{3,0} = - \ln a - \frac{3}{4} - C + b_{1,0};\nonumber\\
&& b_{3,1} = -\Big(\ln a + 1 - b_{1,0}\Big)^2 + b_{1,0} A_1 + \frac{A_2}{4} - D_1 \ln a - D_2.\qquad
\end{eqnarray}

\noindent
In this minimal scheme the $\beta$-function (certainly defined in terms of the renormalized coupling constant) takes the simplest form

\begin{equation}\label{Beta_Minimal}
\widetilde\beta(\alpha) = \frac{\alpha^2 N_f}{\pi}  + \frac{\alpha^3 N_f}{\pi^2} - \frac{\alpha^4 N_f}{2\pi^3} + \frac{\alpha^5 N_f}{2\pi^4} + O(\alpha^6).
\end{equation}

\noindent
Evidently, the minimal scheme is NSVZ. In the considered approximation this can be seen from Eqs. (\ref{Gamma_Minimal}) and (\ref{Beta_Minimal}). In all loops this is also true, because due to the existence of the all-loop NSVZ schemes (constructed, e.g., in \cite{Kataev:2013eta,Kataev:2019olb}) the scheme independent consequences of the NSVZ relation should be satisfied \cite{Kataev:2013csa}.

For completeness, we also present the finite renormalization producing the t'Hooft scheme (in which all terms in the $\beta$-function starting from the three-loop approximation are set to 0) in the considered approximation,

\begin{eqnarray}
&& N_f b_{2,0} = -\frac{1}{2} + N_f\Big(-\ln a -1 -\frac{A_1}{2} + b_{1,0}\Big);\nonumber\\
&& N_f b_{3,0} + (N_f)^2 b_{3,1} = -\frac{1}{2} + N_f\Big(- \ln a - \frac{3}{4} - C + b_{1,0}\Big)\nonumber\\
&&\qquad + (N_f)^2\Big\{ -\Big(\ln a + 1 - b_{1,0}\Big)^2 + b_{1,0} A_1 + \frac{A_2}{4} - D_1 \ln a - D_2\Big\}.\qquad
\end{eqnarray}

\noindent
Certainly, due to the presence of the terms which do not contain $N_f$ this finite renormalization is not compatible with the structure of quantum correction, because any supergraph contributing to the $\beta$-function of the considered theory is evidently proportional to $N_f$. Therefore, the simplest expression for the $\beta$-function in the four-loop approximation is given by Eq. (\ref{Beta_Minimal}).

\subsection{The NSVZ relation for RGFs defined in terms of the renormalized coupling constant}
\hspace*{\parindent}\label{Subsection_NSVZ}

The scheme dependent RGFs defined in terms of the renormalized coupling constant satisfy the NSVZ equation

\begin{equation}\label{NSVZ_Equation_Renormalized}
\widetilde\beta(\alpha) = \frac{\alpha^2 N_f}{\pi}\Big(1-\widetilde\gamma(\alpha)\Big)
\end{equation}

\noindent
only for certain renormalization prescriptions called ``the NSVZ schemes''. According to \cite{Goriachuk:2018cac} (see also \cite{Goriachuk_Conference,Goriachuk:2020wyn,Korneev:2021zdz} for the generalization to the non-Abelian case) these schemes constitute a continuous set and are related by finite renormalizations $\alpha' = \alpha'(\alpha)$, $Z' =z(\alpha) Z$ which satisfy the equation

\begin{equation}\label{NSVZ_Class}
\frac{1}{\alpha'(\alpha)} - \frac{1}{\alpha} - \frac{N_f}{\pi} \ln z(\alpha) = B,
\end{equation}

\noindent
where $B$ is a finite constant.

Comparing the four-loop $\beta$-function (\ref{Beta4_Result}) and the three-loop anomalous dimension (\ref{Gamma3_Result}) we see that the NSVZ equation (\ref{NSVZ_Equation_Renormalized}) is valid if and only if the finite constants $b_i$ and $g_i$ satisfy the relations

\begin{equation}\label{Relations_Between_Constants}
b_{2,0} = g_{1,0};\qquad b_{3,0} = g_{2,0};\qquad b_{3,1} = g_{2,1}.
\end{equation}

\noindent
It is easy to see that these equations follows from Eq. (\ref{NSVZ_Class}). Really, in the HD+MSL scheme all finite constants vanish and RGFs defined in terms of the bare coupling constant coincide with the ones defined in terms of the renormalized coupling constant after the formal replacement $\alpha_0 \to \alpha$. (In the considered approximation this can be verified explicitly by comparing the corresponding expressions.) Certainly, the HD+MSL scheme is NSVZ. This follows from the general result of \cite{Kataev:2013eta} which is based on the statement that RGFs defined in terms of the bare couplings satisfy the NSVZ equation in all loops in the case of using the higher derivative regularization. The scheme determined by Eqs. (\ref{Ln_Z}) and (\ref{Full_Couplings_Relations}) can be obtained from the HD+MSL scheme as a result of the finite renormalization with

\begin{eqnarray}\label{Alpha_Finite_Renormalization}
&& \frac{1}{\alpha'(\alpha)} = \frac{1}{\alpha} + \frac{N_f}{\pi}\, b_{1,0} + \frac{\alpha N_f}{\pi^2}\, b_{2,0} - \frac{\alpha^2 N_f}{2\pi^3} \Big(b_{3,0} + N_f b_{3,1} + 2 N_f b_{1,0}\, b_{2,0}\Big) + O(\alpha^3);\qquad\\
\label{Z_Finite_Renormalization}
&& \ln z = \frac{\alpha}{\pi}\, g_{1,0} - \frac{\alpha^2}{2\pi^2}\Big(g_{2,0} + N_f g_{2,1} + 2 N_f b_{1,0}\, g_{1,0}\Big) + O(\alpha^3).
\end{eqnarray}

\noindent
(Here the coupling constant and the renormalization constant for the matter superfields in the HD+MSL scheme are $\alpha$ and $Z$, respectively.) Note that, for simplicity, we wrote down only the terms which contain finite constants entering the expressions for RGFs in the considered approximation. Substituting these expression into Eq. (\ref{NSVZ_Class}) we obtain the relations (\ref{Relations_Between_Constants}) together with the equation $B = N_f b_{1,0}/\pi$.

\section{Removing the dependence on $N_f$ by a finite renormalization in all loops}
\label{Section_Minimal_Scheme}

\subsection{General statements}
\hspace*{\parindent}\label{Subsection_Minimal_Scheme_General}

In the previous sections we saw that all terms in the anomalous dimension proportional to $(N_f)^k$ with $k\ge 1$ and all terms in the $\beta$-function propotional to $(N_f)^k$ with $k\ge 2$ can be set to 0 by a special choice of a subtraction scheme. Therefore, there is a certain finite renormalization $\alpha\to\alpha'(\alpha)$;\ \ $Z\to Z' = z(\alpha) Z$ which removes these terms. It is very important that the form of this finite renormalization is compatible with the structure of quantum corrections. In (S)QED with $N_f$ flavors this implies that

\begin{equation}
\frac{1}{\alpha'(\alpha)} = \frac{1}{\alpha} + O(N_f);\qquad z(\alpha) = O(1).
\end{equation}

\noindent
(The second condition means that $z(\alpha)$ should not contain terms with negative powers of $N_f$.) Here we will demonstrate that the above statement is valid not only in the considered approximations (three loops for the anomalous dimension and four loops for the $\beta$-function), but also in all orders.

As a starting point we split RGFs into the parts proportional to various powers of $N_f$,

\begin{eqnarray}
&& \widetilde\beta(\alpha) \equiv N_f\, \widetilde\beta_1(\alpha) + (N_f)^2\, \widetilde\beta_2(\alpha) + (N_f)^3\, \widetilde\beta_3(\alpha) + \ldots;\vphantom{\Big(}\qquad\\
&& \widetilde\gamma(\alpha) \equiv \widetilde\gamma_0(\alpha) + N_f\, \widetilde\gamma_1(\alpha) + (N_f)^2\, \widetilde\gamma_2(\alpha) + \ldots \vphantom{\Big(}
\end{eqnarray}

\noindent
The equations which describe how RGFs transform under finite renormalizations were constructed in \cite{Vladimirov:1979ib,Vladimirov:1979my} and are written as

\begin{eqnarray}\label{Beta_Transformation}
&& \widetilde\beta'(\alpha') = \frac{d\alpha'}{d\alpha} \widetilde\beta(\alpha);\\
\label{Gamma_Transformation}
&& \widetilde\gamma'(\alpha') = \frac{d\ln z}{d\alpha} \widetilde\beta(\alpha) + \widetilde\gamma(\alpha).
\end{eqnarray}

\noindent
According to \cite{Kataev:2013csa} the functions $\widetilde\beta_1(\alpha)$ and $\widetilde\gamma_0(\alpha)$ are scheme independent and remain invariant under these finite renormalizations. Moreover, in the supersymmetric case they satisfy the relation

\begin{equation}
\frac{\widetilde\beta_1(\alpha)}{\alpha^2} = \frac{1}{\pi}\Big(1-\widetilde\gamma_0(\alpha)\Big),
\end{equation}

\noindent
which follows from the NSVZ equation and is its scheme independent consequence. Below will construct a renormalization scheme in which all scheme dependent terms in RGFs (defined in terms of the renormalized coupling constant) are set to 0. For the $\beta$-function this implies that

\begin{equation}
\widetilde\beta'(\alpha') = N_f\,\widetilde\beta_1(\alpha').
\end{equation}

\noindent
In this case from the equation (\ref{Beta_Transformation}) we obtain the function $\alpha'(\alpha)$ in the form

\begin{equation}\label{Alpha_General_Equation}
\int \frac{d\alpha'}{\widetilde\beta_1(\alpha')} = N_f \int \frac{d\alpha}{\widetilde\beta(\alpha)}.
\end{equation}

\noindent
However, it is also very important that the difference $1/\alpha'-1/\alpha$ should be proportional to $N_f$, because only in this case the considered finite renormalization is compatible with the structure of quantum corrections. To see this, we note that $\widetilde\beta(\alpha) = N_f \widetilde\beta_1(\alpha) + O(\alpha^4(N_f)^2)$. Therefore, Eq. (\ref{Alpha_General_Equation}) (for the considered theory) can equivalently be rewritten in the form

\begin{eqnarray}\label{Alpha_Auxiliary_Equation}
&& \frac{1}{\pi}\int \frac{d\alpha'}{\widetilde\beta_1(\alpha')} - \frac{1}{\pi}\int \frac{d\alpha}{\widetilde\beta_1(\alpha)} = \frac{1}{\alpha} -\frac{1}{\alpha'} + \frac{N_f}{\pi} b_{1,0} - \frac{1}{\pi}\ln\frac{\alpha'}{\alpha} + \sum\limits_{k=1}^\infty c_k (\alpha'{}^k - \alpha^k) \qquad\nonumber\\
&& = \frac{N_f}{\pi}\int \frac{d\alpha}{\widetilde\beta(\alpha)} - \frac{1}{\pi}\int \frac{d\alpha}{\widetilde\beta_1(\alpha)} = - \frac{N_f}{\pi}\int d\alpha \bigg(\frac{\widetilde\beta_2(\alpha)}{\widetilde\beta_1(\alpha)^2} +\ldots \bigg) = O(\alpha N_f),
\end{eqnarray}

\noindent
where $c_k$ are constants which do not depend on $N_f$. The integration constant in (\ref{Alpha_Auxiliary_Equation}) was denoted by $N_f b_{1,0}/\pi$ in order that Eq. (\ref{Alpha_Finite_Renormalization}) will be valid in the lowest order. From Eq. (\ref{Alpha_Auxiliary_Equation}) it is already clear that the difference $1/\alpha'-1/\alpha$ is proportional to $N_f$, so that the finite renormalization determined by Eq. (\ref{Alpha_General_Equation}) is really compatible with the structure of quantum corrections. In the next section we will illustrate this general statement by an explicit calculation.

Similarly, if we set to 0 all scheme dependent terms in the anomalous dimension of the matter superfields, then

\begin{equation}
\widetilde\gamma'(\alpha') = \widetilde\gamma_0(\alpha'),
\end{equation}

\noindent
and from Eq. (\ref{Gamma_Transformation}) we obtain the required finite renormalization of the matter superfields,

\begin{equation}\label{LnZ_Finite}
\ln z(\alpha) = - \int \frac{d\alpha}{\widetilde\beta(\alpha)} \Big[\,\widetilde\gamma(\alpha) - \widetilde\gamma_0\left(\alpha'(\alpha)\right)\hspace*{-0.6mm}\Big].
\end{equation}

\noindent
It is evident that both the numerator and the denominator of the integrand in the right hand side are proportional to $\alpha^2 N_f$. Taking into account that in the one-loop approximation $\widetilde\beta(\alpha) = \alpha^2 N_f/\pi$ we see that the right hand side of Eq. (\ref{LnZ_Finite}) can be presented as a series in $\alpha$ in which the coefficients do not contain negative powers of $N_f$, so that

\begin{equation}
z(\alpha) = 1 + z_0(\alpha) + N_f z_1(\alpha) + (N_f)^2 z_2(\alpha) + \ldots
\end{equation}

\noindent
Thus, we conclude that using the finite renormalizations compatible with the structure of quantum corrections it is possible to choose such a subtraction scheme in which in all orders

\begin{equation}
\widetilde\gamma'(\alpha)=\widetilde\gamma_0(\alpha);\qquad \widetilde\beta'(\alpha) = N_f\,\widetilde\beta_1(\alpha).
\end{equation}

\subsection{The minimal scheme in the lowest orders}
\hspace*{\parindent}\label{Subsection_Minimal_Scheme_Examples}

Let us illustrate the general argumentation presented in the previous section by an explicit calculation. Namely, we would like to construct such a finite renormalization that transfer the HD+MSL scheme into the minimal scheme and compare the result with Eqs. (\ref{Minimal_Scheme_Original_Constants_G}) and (\ref{Minimal_Scheme_Original_Constants_B}). In the HD+MSL scheme all finite constants $g_i$ and $b_i$ are equal 0, so that RGFs in this scheme are given by the expressions

\begin{eqnarray}\label{Gamma_HD+MSL}
&&\hspace*{-8mm} \widetilde\gamma(\alpha)= - \frac{\alpha}{\pi} + \frac{\alpha^2}{2\pi^2} + \frac{\alpha^2 N_f}{\pi^2}\Big(\ln a + 1 + \frac{A_1}{2}\Big) - \frac{\alpha^3}{2\pi^3}
+ \frac{\alpha^3 N_f}{\pi^3} \Big(- \ln a - \frac{3}{4} - C \Big)  + \frac{\alpha^3 (N_f)^2}{\pi^3} \nonumber\\
&&\hspace*{-8mm} \times \Big\{- (\ln a + 1)^2 + \frac{A_2}{4} - D_1 \ln a - D_2 \Big\} + O(\alpha^4); \vphantom{\frac{N_f^2}{\pi^2}}\\
&& \vphantom{1}\nonumber\\
\label{Beta_HD+MSL}
&&\hspace*{-8mm} \widetilde\beta(\alpha) = \frac{\alpha^2 N_f}{\pi} + \frac{\alpha^3 N_f}{\pi^2} - \frac{\alpha^4 N_f}{2\pi^3} - \frac{\alpha^4 (N_f)^2}{\pi^3}\Big(\ln a + 1+\frac{A_1}{2}\Big) + \frac{\alpha^5 N_f}{2\pi^4} + \frac{\alpha^5 (N_f)^2}{\pi^4} \Big(\ln a + \frac{3}{4}
\nonumber\\
&&\hspace*{-8mm} + C \Big) + \frac{\alpha^5 (N_f)^3}{\pi^4} \Big\{ (\ln a +1)^2  -\frac{A_2}{4} + D_1 \ln a + D_2   \Big\} + O(\alpha^6).
\end{eqnarray}

\noindent
The scheme in which the anomalous dimension and the $\beta$-function are given by Eqs. (\ref{Gamma3_Result}) and (\ref{Beta4_Result}) is obtained from the HD+MSL scheme after the finite renormalization given by Eqs. (\ref{Alpha_Finite_Renormalization}) and (\ref{Z_Finite_Renormalization}). Here our purpose is to find the values of the finite constants $g_i$ and $b_i$ which correspond to the minimal scheme using the method described in Sect. \ref{Subsection_Minimal_Scheme_General}. By other words, the new three-loop anomalous dimension and the new four-loop $\beta$-function should be given by the expressions

\begin{eqnarray}\label{Gamma0}
&& \widetilde\gamma'(\alpha') =  \widetilde\gamma_0(\alpha') = - \frac{\alpha'}{\pi} + \frac{\alpha'{}^2}{2\pi^2} - \frac{\alpha'{}^3}{2\pi^3} + O(\alpha'{}^4);\\
\label{Beta1}
&& \widetilde\beta'(\alpha') = N_f \widetilde\beta_1(\alpha') = \frac{\alpha'{}^2 N_f}{\pi} + \frac{\alpha'{}^3 N_f}{\pi^2} - \frac{\alpha'{}^4 N_f}{2\pi^3} + \frac{\alpha'{}^5 N_f}{2\pi^4} + O(\alpha'{}^6),\qquad
\end{eqnarray}

\noindent
where $\alpha'$ is determined by Eq. (\ref{Alpha_Finite_Renormalization}).

Substituting the expressions (\ref{Beta_HD+MSL}) and (\ref{Beta1}) into Eq. (\ref{Alpha_General_Equation}) and integrating two resulting series we obtain the equation

\begin{eqnarray}
&&\hspace*{-7mm} \frac{N_f}{\pi} b_{1,0} -\frac{1}{\alpha'} - \frac{1}{\pi} \ln \alpha' + \frac{3\alpha'}{2\pi^2} - \frac{5\alpha'{}^2}{4\pi^3} + O(\alpha'{}^3)  = \frac{1}{\pi}\int \frac{d\alpha'}{\widetilde\beta_1(\alpha')} = \frac{N_f}{\pi} \int \frac{d\alpha}{\widetilde\beta(\alpha)} = -\frac{1}{\alpha} - \frac{1}{\pi} \ln\alpha \nonumber\\
&&\hspace*{-7mm} + \frac{3\alpha}{2\pi^2} + \frac{\alpha N_f}{\pi^2}\Big(\ln a + 1 + \frac{A_1}{2}\Big) - \frac{5\alpha^2}{4\pi^3} - \frac{\alpha^2 N_f}{\pi^3}\Big(\frac{3}{2}\ln a + \frac{11}{8} + \frac{C}{2} + \frac{A_1}{2}\Big) - \frac{\alpha^2 (N_f)^2}{2\pi^3}\Big[(\ln a +1)^2 \nonumber\\
&&\hspace*{-7mm} - \frac{A_2}{4} + D_1\ln a + D_2\Big] + O(\alpha^3). \vphantom{\frac{N_f^2}{\pi^2}}
\end{eqnarray}

\noindent
The coupling constant $\alpha'$ in the left hand side should be expressed in terms of $\alpha$ with the help of Eq. (\ref{Alpha_Finite_Renormalization}). After this, equating the coefficients at different powers of alpha we find the values of the coefficients $b_i$,

\begin{eqnarray}
&&\hspace*{-7mm} b_{2,0} = - \ln a -1 - \frac{A_1}{2} + b_{1,0};\nonumber\\
&&\hspace*{-7mm} b_{3,0} = -\ln a - \frac{3}{4} - C + b_{1,0} - 2\Big(b_{2,0} + \ln a +1 +\frac{A_1}{2} - b_{1,0}\Big);\nonumber\\
&&\hspace*{-7mm} b_{3,1} = - (\ln a + 1)^2 + \frac{A_2}{4} - D_1\ln a - D_2 - 2 b_{1,0} b_{2,0} + (b_{1,0})^2,
\end{eqnarray}

\noindent
which coincide with the ones defined by Eq. (\ref{Minimal_Scheme_Original_Constants_B}).

Similarly, substituting the expressions (\ref{Gamma_HD+MSL}), (\ref{Beta_HD+MSL}), and (\ref{Gamma0}) into Eq. (\ref{LnZ_Finite}) after some transformations we construct the expression for $\ln z(\alpha)$ in the considered approximation,

\begin{eqnarray}
&&\hspace*{-7mm} \ln z(\alpha)  = - \int \frac{d\alpha}{\widetilde\beta(\alpha)} \Big[\,\widetilde\gamma(\alpha) - \widetilde\gamma_0\left(\alpha'(\alpha)\right)\hspace*{-0.6mm}\Big]  \nonumber\\
&&\hspace*{-7mm} = - \frac{\alpha}{\pi}\Big(\ln a + 1 + \frac{A_1}{2} - b_{1,0}\Big)
- \frac{\alpha^2}{2\pi^2} \Big(- 2\ln a - \frac{7}{4}  -\frac{A_1}{2} - C - b_{2,0} + 2 b_{1,0}\Big)\nonumber\\
&&\hspace*{-7mm} - \frac{\alpha^2 N_f}{2\pi^2}\Big\{- (\ln a + 1)^2 + \frac{A_2}{4} - D_1 \ln a - D_2 + (b_{1,0})^2\Big\} + O(\alpha^3).
\end{eqnarray}

\noindent
Comparing this expression with Eq. (\ref{Z_Finite_Renormalization}) we obtain the values of the finite constants corresponding to the minimal scheme for the anomalous dimension of the matter superfields,

\begin{eqnarray}
&&\hspace*{-7mm} g_{1,0} = - \ln a - 1 - \frac{A_1}{2} + b_{1,0};\nonumber\\
&&\hspace*{-7mm} g_{2,0} = - 2\ln a - \frac{7}{4} -\frac{A_1}{2} - C - b_{2,0} + 2 b_{1,0};\nonumber\\
&&\hspace*{-7mm} g_{2,1} = - (\ln a + 1)^2 + \frac{A_2}{4} - D_1 \ln a - D_2 + (b_{1,0})^2 - 2b_{1,0}\, g_{1,0}.
\end{eqnarray}

\noindent
Again, it is easy to see that this system of equations is equivalent to Eq. (\ref{Minimal_Scheme_Original_Constants_G}) derived directly from this requirement. Therefore, the general argumentation presented in this section really works and can really reproduce the results in the lowest approximations.

\section{Conclusion}
\hspace*{\parindent}

In this paper we have calculated the three-loop anomalous dimension of the matter superfields and the four-loop $\beta$-function for ${\cal N}=1$ SQED with $N_f$ flavors regularized by higher derivatives. As a starting point we construct the integrals which give the three-loop anomalous dimension defined in terms of the bare coupling constant with the help of a special computer program written by I.S. As a correctness test we have verified the cancellation of the gauge dependence. In the supersymmetric case it takes place because (due to the non-renormalization of the superpotential \cite{Grisaru:1979wc}) the anomalous dimension of the chiral matter superfields is related to the anomalous dimension of their mass. The obtained integrals have been calculated using the Chebyshev polynomial method \cite{Rosner:1967zz}. The result for the scheme independent part of the three-loop contribution (which does not contain $N_f$) coincided with the one obtained with the use of dimensional reduction \cite{Jack:1996qq}. The parts proportional to $N_f$ and $(N_f)^2$ turn out to depend on regularization parameters. Some of them can be calculated analytically for the simplest regulator functions, but one of them (in the $(N_f)^2$ contribution) can be obtained only numerically.

The four-loop $\beta$-function defined in terms of the bare coupling constant can immediately be found using the general statement \cite{Stepanyantz:2011jy,Stepanyantz:2014ima} that the NSVZ equation is valid for RGFs defined in terms of the bare coupling constant in all orders in the case of using the higher derivative regularization. (Due to the scheme independence of these RGFs this is true for an arbitrary renormalization prescription supplementing this regularization.)

The three-loop anomalous dimension of the matter superfields and the four-loop $\beta$-function defined in terms of the renormalized coupling constant have been obtained from the corresponding RGFs defined in terms of the bare coupling constant in an arbitrary subtraction scheme. For a certain renormalization prescription the results coincided with the ones in $\overline{\mbox{DR}}$ scheme. Also it turned out that by a special choice of a renormalization prescription compatible with the structure of quantum corrections one can set to 0 all terms proportional to $(N_f)^k$ with $k\ge 1$ in the anomalous dimension and all terms proportional to $(N_f)^k$ with $k\ge 2$ in the $\beta$-function. We have demonstrated that this can be done in all orders. Note that the proof is also valid for other Abelian gauge theories, say, for usual QED with $N_f$ flavors. In this ``minimal'' scheme RGFs defined in terms of the renormalized couplings take the simplest form. For ${\cal N}=1$ SQED this minimal scheme is NSVZ in all orders, and only scheme independent terms survive in it.

We believe that it would be also interesting to generalize the results of this paper to the non-Abelian case. Possibly, they will be also useful for investigating fixed point in supersymmetric theories, see, e.g., \cite{Seiberg:1994pq,Ryttov:2017khg,Bond:2022xvr} and references therein.

\section*{Acknowledgments}
\hspace*{\parindent}

K.S. would like to express the gratitude to Prof. M.A.Shifman for valuable discussions. I.S. is very grateful to S.V.Morozov for valuable advices about algorithms and for the tutorial about C++ language capabilities.

The work was supported by Foundation for Advancement of Theoretical Physics and Mathematics ``BASIS'', grants  No. 19-1-1-45-3 (I.S.) and 19-1-1-45-1 (K.S.).

\appendix

\section{Integrals needed for calculating the three-loop anomalous dimension}
\label{Appendix_Integrals_For_Gamma}

\subsection{Integrals which do not contain $N_f$}
\hspace*{\parindent}

First, let us calculate a part of the three-loop contribution to the anomalous dimension of the matter superfields which does not contain $N_f$. It is important that all terms coming from the renormalization of the coupling constant in the previous orders are proportional to $N_f$ or $(N_f)^2$. Therefore, the considered contribution can be written in the form

\begin{eqnarray}
&& I_0 \equiv e^6\, \frac{d}{d\ln\Lambda} \int \frac{d^4K}{(2\pi)^4} \frac{d^4L}{(2\pi)^4} \frac{d^4Q}{(2\pi)^4} \frac{8}{R_K R_L R_Q}\bigg[- \frac{1}{3K^4 L^4 Q^4} + \frac{1}{K^4 L^2 Q^4 (Q+L)^2} \qquad\nonumber\\
&& + \frac{1}{K^2 L^4 (K+L)^2 (Q+K+L)^2}\bigg(\frac{1}{Q^2} - \frac{2}{(Q+L)^2}\bigg) \bigg],
\end{eqnarray}

\noindent
where we took into account that the difference between $e_0^6$ and $e^6$ is of the order $e^8$ and can be ignored.

Making in the two parts of the last term the changes of integration variables $K_\mu\to K_\mu-L_\mu$; $L_\mu\to - L_\mu$ and $Q_\mu\to Q_\mu-L_\mu$ we can present this term in the form

\begin{equation}
e^6\, \frac{d}{d\ln\Lambda}\int \frac{d^4K}{(2\pi)^4} \frac{d^4L}{(2\pi)^4} \frac{d^4Q}{(2\pi)^4} \frac{8}{K^2 L^4 Q^2 (K+L)^2 (Q+K)^2} \bigg(\frac{1}{R_{K+L} R_L R_Q} - \frac{2}{R_K R_L R_{Q-L}}\bigg).
\end{equation}

\noindent
To simplify this expression, we note that

\begin{eqnarray}\label{Auxiliary_Identity1}
&& \frac{d}{d\ln\Lambda}\int \frac{d^4K}{(2\pi)^4} \frac{d^4L}{(2\pi)^4} \frac{d^4Q}{(2\pi)^4} \frac{1}{K^2 L^4 Q^2 (K+L)^2 (Q+K)^2 R_L R_Q} \bigg(\frac{1}{R_{K+L}} - \frac{1}{R_K}\bigg) = 0;\qquad\\
\label{Auxiliary_Identity2}
&& \frac{d}{d\ln\Lambda}\int \frac{d^4K}{(2\pi)^4} \frac{d^4L}{(2\pi)^4} \frac{d^4Q}{(2\pi)^4} \frac{1}{K^2 L^4 Q^2 (K+L)^2 (Q+K)^2 R_K R_L} \bigg(\frac{1}{R_{Q}} - \frac{1}{R_{Q-L}}\bigg) = 0
\end{eqnarray}

\noindent
because both integrals (without the derivative with respect to $\ln\Lambda$) are finite in both infrared and ultraviolet regions. This implies that they are finite constants which do not depend on $\Lambda$ and, therefore, vanish after the differentiation with respect to $\ln\Lambda$.

With the help of Eqs. (\ref{Auxiliary_Identity1}) and (\ref{Auxiliary_Identity2}) the considered contribution to the anomalous dimension can be presented in the form

\begin{eqnarray}\label{I0_Expression}
&& I_0 \equiv 8e^6\, \int \frac{d^4K}{(2\pi)^4} \frac{d^4L}{(2\pi)^4} \frac{d^4Q}{(2\pi)^4} \frac{d}{d\ln\Lambda}\bigg(\frac{1}{R_K R_L R_Q}\bigg)\bigg( - \frac{1}{3K^4 L^4 Q^4} + \frac{1}{K^4 L^2 Q^4 (Q+L)^2} \qquad\nonumber\\
&& - \frac{1}{K^2 L^4 Q^2 (K+L)^2 (Q+K)^2} \bigg).
\end{eqnarray}

In the first term the angle integration is trivial, so that it can equivalently be rewritten as

\begin{equation}\label{First_Term_Original}
-\frac{8e^6}{3(8\pi^2)^3}\int\limits_0^\infty \frac{dK}{K} \int\limits_0^\infty \frac{dL}{L} \int\limits_0^\infty \frac{dQ}{Q} \frac{d}{d\ln\Lambda}\bigg(\frac{1}{R_K R_L R_Q}\bigg).
\end{equation}

\noindent
It is convenient to break the integration domain into 6 parts defined by the conditions $Q>K>L$,\ \ $Q>L>K$,\ \ $K>Q>L$,\ \ $K>L>Q$,\ \ $L>K>Q$,\ \ $L>Q>K$. Evidently, due to the symmetry of the integrand integrations over all these domains give the same results, so that the expression (\ref{First_Term_Original}) can be presented as

\begin{equation}\label{First_Term}
-\frac{2\alpha^3}{\pi^3}\int\limits_0^\infty \frac{dK}{K} \int\limits_0^K \frac{dL}{L} \int\limits_0^L \frac{dQ}{Q} \frac{d}{d\ln\Lambda}\bigg(\frac{1}{R_K R_L R_Q}\bigg).
\end{equation}

The second term in the expression (\ref{I0_Expression}) can be calculated using the technique of the Chebyshev polynomials. Namely, using Eqs. (\ref{Inverse_Square}) (for $(Q+L)^{-2}$) and (\ref{Ortogonality}) it can be presented as

\begin{eqnarray}\label{Second_Term_Original}
&& 8e^6\, \int \frac{d^4K}{(2\pi)^4} \frac{d^4L}{(2\pi)^4} \frac{d^4Q}{(2\pi)^4} \frac{d}{d\ln\Lambda}\bigg(\frac{1}{R_K R_L R_Q}\bigg) \frac{1}{K^4 L^2 Q^4 (Q+L)^2}\nonumber\\
&& = \frac{\alpha^3}{\pi^3} \int\limits_0^\infty \frac{dK}{K} \int\limits_0^\infty \frac{dQ}{Q} \bigg(\int\limits_0^Q \frac{dL\, L}{Q^2} + \int\limits_Q^\infty \frac{dL}{L}\bigg)\frac{d}{d\ln\Lambda}\bigg(\frac{1}{R_K R_L R_Q}\bigg).
\end{eqnarray}

\noindent
In the first term $Q>L$ and, therefore, the integration domain consists of three parts: $K>Q>L$,\ \ $Q>K>L$, and $Q>L>K$. Similarly, in the second term $L>Q$ and the integration domain consists of the parts $K>L>Q$,\ \ $L>K>Q$, and $L>Q>K$. Breaking the integral into the sum of integrals over these domains, changing the order of integrations, and renaming the integration variables we can be rewrite the expression (\ref{Second_Term_Original}) in the form

\begin{equation}\label{Second_Term}
\frac{\alpha^3}{\pi^3} \int\limits_0^\infty dK \int\limits_0^K dL \int\limits_0^L dQ \bigg(\frac{Q}{KL^3} + \frac{Q}{LK^3} + \frac{L}{QK^3} + \frac{3}{KLQ}\bigg) \frac{d}{d\ln\Lambda}\bigg(\frac{1}{R_K R_L R_Q}\bigg).
\end{equation}

The third term in the expression (\ref{I0_Expression}) can also be calculated with the help of Chebyshev polynomilas. Using Eqs. (\ref{Inverse_Square}), (\ref{Product}), and (\ref{Ortogonality}) it can be presented in the form

\begin{eqnarray}\label{Third_Term}
&&\hspace*{-7mm} - 8e^6\, \int \frac{d^4K}{(2\pi)^4} \frac{d^4L}{(2\pi)^4} \frac{d^4Q}{(2\pi)^4} \frac{d}{d\ln\Lambda}\bigg(\frac{1}{R_K R_L R_Q}\bigg) \frac{1}{K^2 L^4 Q^2 (K+L)^2 (Q+K)^2}\nonumber\\
&&\hspace*{-7mm} = - \frac{\alpha^3}{\pi^3} \int\limits_0^\infty \frac{dK}{K} \bigg(\int\limits_0^K \frac{dL}{L K^2} + \int\limits_K^\infty \frac{dL}{L^3} \bigg)
\bigg(\int\limits_0^K dQ\, Q + \int\limits_K^\infty \frac{dQ\, K^2}{Q}\bigg)\frac{d}{d\ln\Lambda}\bigg(\frac{1}{R_K R_L R_Q}\bigg)\nonumber\\
&&\hspace*{-7mm} = - \frac{\alpha^3}{\pi^3} \int\limits_0^\infty dK \int\limits_0^K dL \int\limits_0^L dQ \bigg(\frac{3Q}{LK^3} + \frac{L}{QK^3} +\frac{1}{KLQ} + \frac{Q}{KL^3}\bigg)\frac{d}{d\ln\Lambda}\bigg(\frac{1}{R_K R_L R_Q}\bigg).
\end{eqnarray}

Summing up the integrals (\ref{First_Term}), (\ref{Second_Term}), and (\ref{Third_Term}) we obtain that the considered contribution to the three-loop anomalous dimension is given by the expression

\begin{equation}
I_0 = -\frac{2\alpha^3}{\pi^3} \int\limits_0^\infty dK \int\limits_0^K dL \int\limits_0^L dQ\, \frac{d}{d\ln\Lambda}\bigg(\frac{1}{R_K R_L R_Q}\bigg)\, \frac{Q}{LK^3}.
\end{equation}

\noindent
After the change of the integration variable $Q = x L$ it can be rewritten as

\begin{equation}
I_0 = - \frac{2 \alpha^3}{\pi^3} \int\limits_0^\infty dK \int\limits_0^K dL \int\limits_0^1 dx\, \frac{d}{d\ln\Lambda}\bigg(\frac{1}{R_K R_L R_{xL}}\bigg)\, \frac{xL}{K^3}.
\end{equation}

\noindent
In this expression we again make the change of the integration variable $L = y K$. The result takes the form

\begin{equation}
I_0 = - \frac{2 \alpha^3}{\pi^3} \int\limits_0^\infty dK \int\limits_0^1 dy \int\limits_0^1 dx\, \frac{d}{d\ln\Lambda}\bigg(\frac{1}{R_K R_{yK} R_{xyK}}\bigg)\, \frac{xy}{K}.
\end{equation}

\noindent
All regulator functions $R$ present in this expression depend on $K$ and $\Lambda$ only in the combination $K/\Lambda$. Therefore, the derivative with respect to $\ln\Lambda$ can be expressed in terms of the derivative with respect to $\ln K$,

\begin{equation}
\frac{d}{d\ln\Lambda}\bigg(\frac{1}{R_K R_{yK} R_{xyK}}\bigg) = - \frac{d}{d\ln K}\bigg(\frac{1}{R_K R_{yK} R_{xyK}}\bigg),
\end{equation}

\noindent
and the resulting integral over $K$ turns out to be the integral of a total derivative,

\begin{equation}
I_0 = \frac{2 \alpha^3}{\pi^3} \int\limits_0^\infty dK \int\limits_0^1 dy \int\limits_0^1 dx\, \frac{d}{dK}\bigg(\frac{1}{R_K R_{yK} R_{xyK}}\bigg)\, xy.
\end{equation}

\noindent
Taking into account that the regulator function $R(x)$ is equal to 1 at $x=0$ and rapidly increases at infinity, we obtain the resulting expression for a part of the three-loop anomalous dimension which does not depend on $N_f$,

\begin{equation}\label{I0_Result}
I_0 = - \frac{2 \alpha^3}{\pi^3} \int\limits_0^1 dy \int\limits_0^1 dx\, xy = - \frac{\alpha^3}{2\pi^3}.
\end{equation}

\noindent
According to \cite{Kataev:2013csa}, this contribution is scheme independent and, therefore, should coincide with the one obtained in the $\overline{\mbox{DR}}$ scheme. This coincidence really takes place and confirmes the correctness of the calculation.

\subsection{Integrals linear in $N_f$ containing an insertion of the one-loop polarization operator}
\hspace*{\parindent}

Next, let us consider a part of the contribution proportional to $(N_f)^1$ which comes from the supergraphs containing an insertion of the one-loop polarization operator. Note that in this case it is necessary to take into account the renormalization of the coupling constant in the one-loop approximation. The corresponding contribution is obtained when in the two-loop contribution the bare coupling constant is expressed in terms of the renormalized one with the help of Eq. (\ref{Relation_Between_Couplings}). Then the resulting expression for considered part of the three-loop anomalous dimension can be written as

\begin{eqnarray}\label{With_One_1Loop_Polarization}
&& I_{1,1} \equiv 16 N_f e^6 \frac{d}{d\ln\Lambda} \int \frac{d^4K}{(2\pi)^4} \frac{d^4L}{(2\pi)^4} \frac{K_\mu L^\mu}{R_K^2 R_L K^4 L^4 (K+L)^2}
\bigg\{\int\frac{d^4Q}{(2\pi)^4}\bigg(\frac{1}{Q^2(Q+K)^2} \qquad\nonumber\\
&& - \frac{1}{(Q^2+M^2)((Q+K)^2 + M^2)}\bigg) - \frac{1}{8\pi^2} R_K \Big(\ln\frac{\Lambda}{\mu} + b_{1,0}\Big)\bigg\}.
\end{eqnarray}

\noindent
It is convenient to introduce the function

\begin{eqnarray}\label{P_Definition}
&& p(K) \equiv \int\frac{d^4Q}{(2\pi)^4}\bigg(\frac{1}{Q^2(Q+K)^2} - \frac{1}{(Q^2+M^2)((Q+K)^2 + M^2)}\bigg) - \frac{1}{8\pi^2} R_K \ln\frac{\Lambda}{K} \qquad\nonumber\\
&& = \frac{1}{8\pi^2} \bigg\{(1-R_K) \ln\frac{\Lambda}{K} + \ln a + \sqrt{1+\frac{4M^2}{K^2}} \mbox{arctanh} \sqrt{\frac{K^2}{K^2+4M^2}}\,\bigg\}.\qquad
\end{eqnarray}

\noindent
From the explicit form of the function $p(K)$ it s easy to see that

\begin{equation}\label{P0}
p(0) = \frac{1}{8\pi^2}\Big(\ln a + 1\Big).
\end{equation}

\noindent
Let us rewrite the expression (\ref{With_One_1Loop_Polarization}) in the form

\begin{eqnarray}\label{I11_Equivalent}
&& I_{1,1} = 16 N_f e^6 \frac{d}{d\ln\Lambda} \int \frac{d^4K}{(2\pi)^4} \frac{d^4L}{(2\pi)^4} \frac{K_\mu L^\mu}{R_K^2 R_L K^4 L^4 (K+L)^2}\, p(K) \nonumber\\
&& + \frac{2 N_f e^6}{\pi^2} \int \frac{d^4K}{(2\pi)^4} \frac{d^4L}{(2\pi)^4} \frac{K_\mu L^\mu}{K^4 L^4 (K+L)^2} \frac{d}{d\ln\Lambda}\Big(\frac{1}{R_K R_L}\Big) \ln\frac{\Lambda}{K} \nonumber\\
&& - \frac{2 N_f e^6}{\pi^2} \Big(\ln\frac{\Lambda}{\mu} + b_{1,0}\Big) \frac{d}{d\ln\Lambda} \int \frac{d^4K}{(2\pi)^4} \frac{d^4L}{(2\pi)^4} \frac{K_\mu L^\mu}{R_K R_L K^4 L^4 (K+L)^2}\qquad
\end{eqnarray}

\noindent
and calculate the integrals entering it.

First, using the technique of Chebyshev polynomials we present the integral

\begin{eqnarray}
&& I_{1,1,1}\equiv \frac{2 N_f e^6}{\pi^2} \frac{d}{d\ln\Lambda} \int \frac{d^4K}{(2\pi)^4} \frac{d^4L}{(2\pi)^4} \frac{K_\mu L^\mu}{R_K R_L K^4 L^4 (K+L)^2}\nonumber\\
&& =  \frac{2 N_f e^6}{\pi^2} \int \frac{d^4K}{(2\pi)^4} \frac{d^4L}{(2\pi)^4} \frac{d}{d\ln\Lambda}\Big(\frac{1}{R_K R_L}\Big) \bigg( \frac{1}{2K^4 L^4} - \frac{1}{K^2 L^4 (K+L)^2} \bigg)
\end{eqnarray}

\noindent
in the form

\begin{equation}
I_{1,1,1} = - \frac{2 N_f \alpha^3}{\pi^3}  \int\limits_0^\infty dK \int\limits_K^\infty dL\, \frac{K}{L^3} \frac{d}{d\ln\Lambda}\Big(\frac{1}{R_K R_L}\Big).
\end{equation}

\noindent
Making the change of the integration variable $L \equiv x K$ and taking into account that the function $(R_K R_{xK})^{-1}$ depends on $K$ and $\Lambda$ only in the combination $K/\Lambda$ we can calculate this integral for an arbitrary regulator function $R$,

\begin{equation}\label{I111_Result}
I_{1,1,1} = \frac{2 N_f \alpha^3}{\pi^3}  \int\limits_0^\infty dK \int\limits_1^\infty dx\, \frac{1}{x^3} \frac{d}{dK}\Big(\frac{1}{R_K R_{xK}}\Big) = - \frac{2 N_f \alpha^3}{\pi^3} \int\limits_1^\infty \frac{dx}{x^3}
= - \frac{N_f\alpha^3}{\pi^3}.
\end{equation}

Next, we consider the integral

\begin{equation}
I_{1,1,2} \equiv 16 N_f e^6 \frac{d}{d\ln\Lambda} \int \frac{d^4K}{(2\pi)^4} \frac{d^4L}{(2\pi)^4} \frac{K_\mu L^\mu}{R_K^2 R_L K^4 L^4 (K+L)^2}\, p(K)
\end{equation}

\noindent
and calculate its angular part using the Chebyshev polynomials. Then, after some transformations, it can be rewritten in the form

\begin{equation}
I_{1,1,2} = - \frac{2 N_f e^6}{(4\pi^2)^2} \frac{d}{d\ln\Lambda} \int\limits_0^\infty dK \int\limits_0^K dL \frac{1}{R_K R_L} \Big(\frac{L\, p(K)}{R_K K^3} + \frac{L\, p(L)}{R_L K^3}\Big).
\end{equation}

\noindent
After the change of the integration variable $L \equiv x K$ this integral can be reduced to the integral of the total derivative with respect to $K$,

\begin{eqnarray}
&& I_{1,1,2} = -\frac{8 N_f\alpha^3}{\pi} \frac{d}{d\ln\Lambda} \int\limits_0^1 dx \int\limits_0^\infty dK \frac{1}{R_K R_{xK}} \Big(\frac{x p(K)}{R_K K} + \frac{x p(xK)}{R_{xK} K}\Big) \qquad\nonumber\\
&& = \frac{8 N_f\alpha^3}{\pi}  \int\limits_0^1 dx \int\limits_0^\infty dK \frac{d}{dK} \Big(\frac{x\, p(K)}{R_K^2 R_{xK}} + \frac{x\, p(xK)}{R_K R_{xK}^2}\Big).
\end{eqnarray}

\noindent
Calculating it and taking Eq. (\ref{P0}) into account we obtain

\begin{equation}\label{I112_Result}
I_{1,1,2} = - \frac{16 N_f\alpha^3}{\pi}  \int\limits_0^1 dx\, x\, p(0) = - \frac{N_f\alpha^3}{\pi^3}\Big(\ln a+1\Big).
\end{equation}

The remaining integral in the expression (\ref{With_One_1Loop_Polarization}),

\begin{equation}
I_{1,1,3} \equiv \frac{2 N_f e^6}{\pi^2}  \int \frac{d^4K}{(2\pi)^4} \frac{d^4L}{(2\pi)^4} \frac{K_\mu L^\mu}{K^4 L^4 (K+L)^2}\,\frac{d}{d\ln\Lambda}\Big(\frac{1}{R_K R_L}\Big) \ln\frac{\Lambda}{K},
\end{equation}

\noindent
after performing integration by angles with the help of the Chebyshev polynomials takes the form

\begin{eqnarray}
&& I_{1,1,3} = - \frac{N_f \alpha^3}{\pi^3} \int\limits_0^\infty dK \int\limits_0^K dL\, \frac{d}{d\ln\Lambda} \Big(\frac{1}{R_K R_L}\Big) \frac{L}{K^3} \Big( \ln\frac{\Lambda}{K} + \ln\frac{\Lambda}{L}\Big)\nonumber\\
&& = \frac{N_f \alpha^3}{\pi^3} \int\limits_0^1 dx\, x \int\limits_0^\infty dK \frac{d}{dK} \Big(\frac{1}{R_K R_{xK}}\Big) \Big(2 \ln\frac{\Lambda}{K} -\ln x\Big).
\end{eqnarray}

\noindent
The part of this expression containing $\ln x$ can be calculated for an arbitrary regulator function $R$,

\begin{equation}
- \frac{N_f \alpha^3}{\pi^3} \int\limits_0^1 dx\,x\,\ln x \int\limits_0^\infty dK \frac{d}{dK} \Big(\frac{1}{R_K R_{xK}}\Big)  = \frac{N_f \alpha^3}{\pi^3} \int\limits_0^1 dx\,x\,\ln x = - \frac{N_f \alpha^3}{4\pi^3}.
\end{equation}

\noindent
However, unfortunately, we did not manage to do this for the remaining contribution. After the change of the integration variable $y=K^2/\Lambda^2$ it can be reduced to the expression proportional to the constant $C$ defined by Eq. (\ref{Constants_Definition}),

\begin{equation}\label{I113_Result}
I_{1,1,3} = - \frac{N_f\alpha^3}{\pi^3} \bigg(\int\limits_0^1 dx \int\limits_0^\infty dy\,x\,\ln y\, \frac{d}{dy}\Big(\frac{1}{R(y) R(x^2 y)}\Big)  +\frac{1}{4}\bigg)
= - \frac{N_f\alpha^3}{\pi^3}\Big(C+\frac{1}{4}\Big).
\end{equation}

\noindent
For the regulator function $R(x) = 1+x^n$ the expression for this constant can be expressed in terms of the logarithmic derivative of the gamma-function. For this purpose we first calculate the integral over $y$,

\begin{eqnarray}
C\Big|_{R(x) = 1+ x^n} = \int\limits_0^1 dx \int\limits_0^\infty dy\, x\, \ln y\, \frac{d}{dy}\bigg(\frac{1}{(1+y^n) (1+x^{2n}y^{n})}\bigg) = 2\int\limits_0^1 dx\, x^{2n+1}\, \frac{\ln x}{x^{2n}-1},
\end{eqnarray}

\noindent
and then make the substitution $t=x^{2n}$, after which the considered expression takes the form

\begin{equation}
C\Big|_{R(x) = 1+ x^n} = - \frac{1}{2n^2} \int\limits_0^1 dt\, t^{1/n}\, \frac{\ln t}{1-t} = \frac{1}{2n^2} \psi'\Big(\frac{n+1}{n}\Big).
\end{equation}

\noindent
For $n=1$ and $n=2$ this expression is given by Eq. (\ref{C_Values}).

Substituting the expressions (\ref{I111_Result}), (\ref{I112_Result}), and (\ref{I113_Result}) into Eq. (\ref{I11_Equivalent}) we obtain the result for the considered contribution to the three-loop anomalous dimension

\begin{equation}\label{I11_Result}
I_{1,1} = -\Big(\ln\frac{\Lambda}{\mu} + b_{1,0}\Big) I_{1,1,1} + I_{1,1,2} + I_{1,1,3} = \frac{N_f\alpha^3}{\pi^3} \Big(\ln\frac{\Lambda}{\mu} + b_{1,0}\Big)- \frac{N_f\alpha^3}{\pi^3}\Big(\ln a + C +\frac{5}{4}\Big).
\end{equation}

\subsection{Integrals linear in $N_f$ containing an insertion of the two-loop polarization operator}
\hspace*{\parindent}

The second part of the contribution to the three-loop anomalous dimension proportional to $(N_f)^1$ comes from superdiagrams containing an insertion of the two-loop polarization operator of the quantum gauge superfield. For calculating it we should also take into account the corresponding term in the renormalization of the coupling constant. Then the expression for the considered contribution can be written as

\begin{eqnarray}\label{With_One_2Loop_Polarization}
&&\hspace*{-5mm} I_{1,2} \equiv 8 N_f e^6 \frac{d}{d\ln\Lambda} \int \frac{d^4K}{(2\pi)^4} \frac{1}{R_K^2 K^4}\bigg\{\int \frac{d^4L}{(2\pi)^4} \frac{d^4Q}{(2\pi)^4} \frac{1}{R_L L^2}\bigg(\, \frac{2(Q+K+L)^2-K^2-L^2}{Q^2 (Q+K)^2 (Q+L)^2}\nonumber\\
&&\hspace*{-5mm} \times \frac{1}{(Q+K+L)^2} - \frac{2(Q+K+L)^2-K^2-L^2}{(Q^2+M^2) ((Q+K)^2+M^2) ((Q+L)^2+M^2) ((Q+K+L)^2+M^2)}\nonumber\\
&&\hspace*{-5mm} + \frac{4M^2}{(Q^2+M^2)^2 ((Q+K)^2+M^2) ((Q+L)^2+M^2)}\bigg)  - \frac{1}{(8\pi^2)^2} R_K \Big(\ln\frac{\Lambda}{\mu} + b_{2,0}\Big)\bigg\}.
\end{eqnarray}

\noindent
It is convenient to introduce the function

\begin{eqnarray}
&&\hspace*{-5mm} q(K) \equiv \int \frac{d^4L}{(2\pi)^4} \frac{d^4Q}{(2\pi)^4} \frac{1}{R_L L^2}\bigg(\, \frac{2(Q+K+L)^2-K^2-L^2}{Q^2 (Q+K)^2 (Q+L)^2 (Q+K+L)^2}\nonumber\\
&&\hspace*{-5mm} - \frac{2(Q+K+L)^2-K^2-L^2}{(Q^2+M^2) ((Q+K)^2+M^2) ((Q+L)^2+M^2) ((Q+K+L)^2+M^2)}\nonumber\\
&&\hspace*{-5mm} + \frac{4M^2}{(Q^2+M^2)^2 ((Q+K)^2+M^2) ((Q+L)^2+M^2)}\bigg) - \frac{1}{(8\pi^2)^2} R_K \ln \frac{\Lambda}{K}.
\end{eqnarray}

\noindent
Then the expression (\ref{With_One_2Loop_Polarization}) can equivalently be presented in the form

\begin{eqnarray}
&&\hspace*{-5mm} I_{1,2} = - \frac{8N_f e^6}{(8\pi^2)^2} \Big(\ln\frac{\Lambda}{\mu} + b_{2,0}\Big) \frac{d}{d\ln\Lambda} \int \frac{d^4K}{(2\pi)^4} \frac{1}{R_K K^4}
+ 8N_f e^6 \frac{d}{d\ln\Lambda} \int \frac{d^4K}{(2\pi)^4} \frac{q(K)}{R_K^2 K^4} \nonumber\\
&&\hspace*{-5mm} + \frac{8 N_f e^6}{(8\pi^2)^2} \int \frac{d^4K}{(2\pi)^4} \frac{1}{K^4}\frac{d}{d\ln\Lambda}\Big(\frac{1}{R_K}\Big) \ln\frac{\Lambda}{K}.
\end{eqnarray}

\noindent
To calculate the integrals which enter it, we use the identity

\begin{equation}\label{General_Integral}
\frac{d}{d\ln\Lambda} \int \frac{d^4K}{(2\pi)^4} \frac{f(K/\Lambda)}{K^4} = - \frac{1}{8\pi^2} \int\limits_0^\infty dK\, \frac{df(K/\Lambda)}{dK} = \frac{1}{8\pi^2} f(0)
\end{equation}

\noindent
valid for a nonsingular function $f(K/\Lambda)$ which rapidly decreases at infinity. With the help of Eq. (\ref{General_Integral}) we immediately obtain

\begin{eqnarray}
&& I_{1,2,1} \equiv \frac{8N_f e^6}{(8\pi^2)^2} \frac{d}{d\ln\Lambda} \int \frac{d^4K}{(2\pi)^4} \frac{1}{R_K K^4} = \frac{8N_f e^6}{(8\pi^2)^3} = \frac{N_f \alpha^3}{\pi^3};\qquad \nonumber\\
&& I_{1,2,2} \equiv 8N_f e^6 \frac{d}{d\ln\Lambda} \int \frac{d^4K}{(2\pi)^4} \frac{q(K)}{R_K^2 K^4} = \frac{N_f e^6}{\pi^2} q(0) = \frac{N_f \alpha^3}{\pi^3} (8\pi^2)^2 q(0).
\end{eqnarray}

\noindent
The remaining integral

\begin{eqnarray}
I_{1,2,3} \equiv \frac{8 N_f e^6}{(8\pi^2)^2} \int \frac{d^4K}{(2\pi)^4} \frac{1}{K^4} \frac{d}{d\ln\Lambda}\Big(\frac{1}{R_K}\Big) \ln\frac{\Lambda}{K} = \frac{4 N_f e^6}{(8\pi^2)^3} \int\limits_0^\infty \frac{dK}{K}\,\frac{d}{d\ln K} \Big(\frac{1}{R_K}\Big) \ln\frac{K^2}{\Lambda^2}
\end{eqnarray}

\noindent
is reduced to the constant $A_1$ (defined by Eq. (\ref{Constants_Definition})) after the change of the integration variable $x = K^2/\Lambda^2$,

\begin{equation}
I_{1,2,3} = \frac{N_f \alpha^3}{2\pi^3} \int\limits_0^\infty dx\,\ln x\, \frac{d}{dx}\Big(\frac{1}{R(x)}\Big) = \frac{N_f\alpha^3}{2\pi^3} A_1.
\end{equation}

\noindent
For the regulator function $R(x) = 1+x^n$ this constant vanishes.

Therefore, it remains only to calculate the constant $q(0)$. For this purpose we first present $q(K)$ as a sum

\begin{eqnarray}
q(K) = q_1(K) + q_2(K)
\end{eqnarray}

\noindent
where the functions $q_1(K)$ and $q_2(K)$ are defined by the equations

\begin{eqnarray}
&&\hspace*{-5mm} q_1(K) \equiv \int \frac{d^4L}{(2\pi)^4} \frac{d^4Q}{(2\pi)^4} \frac{1}{R_L L^2}\bigg(\, \frac{2(Q+K+L)^2-K^2-L^2}{Q^2 (Q+K)^2 (Q+L)^2 (Q+K+L)^2}\nonumber\\
&&\hspace*{-5mm} - \frac{2(Q+K+L)^2-K^2-L^2}{(Q^2+M^2) ((Q+K)^2+M^2) ((Q+L)^2+M^2) ((Q+K+L)^2+M^2)} \\
&&\hspace*{-5mm} + \frac{4M^2}{(Q^2+M^2)^2 ((Q+K)^2+M^2) ((Q+L)^2+M^2)}  - \frac{2L_\mu (Q+L)^\mu}{L^2 Q^2 (Q+L)^2 (Q+K+L)^2} \bigg);\nonumber\\
&& \vphantom{1}\nonumber\\
&&\hspace*{-5mm} q_2(K) \equiv \int \frac{d^4L}{(2\pi)^4} \frac{d^4Q}{(2\pi)^4} \frac{2L_\mu (Q+L)^\mu}{R_L L^4 Q^2 (Q+L)^2 (Q+K+L)^2} - \frac{1}{(8\pi^2)^2} R_K \ln \frac{\Lambda}{K}.
\end{eqnarray}

\noindent
Taking the limit $K\to 0$ in the expression for the function $q_1(K)$ we obtain that the value $q_1(0)$ can be written as a vanishing integral of a total derivative in the momentum space,

\begin{eqnarray}
&& q_1(0) = \int \frac{d^4L}{(2\pi)^4} \frac{d^4Q}{(2\pi)^4} \bigg\{\frac{\partial}{\partial Q_\mu}\bigg(-\frac{Q_\mu}{R_L L^4 Q^2 (Q+L)^2} - \frac{Q_\mu}{R_L L^2 Q^4 (Q+L)^2}\bigg)\qquad\nonumber\\
&& - \frac{1}{4}\frac{\partial^2}{\partial Q_\mu \partial Q^\mu} \bigg(\frac{1}{R_L L^2 (Q^2+M^2)((Q+L)^2+M^2)}\bigg)\bigg\} = 0.
\end{eqnarray}

\noindent
The term with the Pauli--Villars masses vanishes due to the absence of singular contributions, while the first two terms cancel each other. (The first one is reduced to the integral over an infinitely large sphere $S^3_\infty$, while the second one is reduced to the surface integral over an infinitely small sphere $S^3_\varepsilon$ surrounding the point $Q_\mu=0$.)

Thus, to obtain the value $q(0)$, it remains to calculate $q_2(0)$. After the changes of the integration variables $Q_\mu \to Q_\mu- L_\mu$,\ \ $L_\mu \to - L_\mu$ this expression can be presented in the form

\begin{eqnarray}
&& q(0) = q_2(0) = \lim\limits_{K\to 0}\bigg\{- \frac{1}{(8\pi^2)^2}\ln\frac{\Lambda}{K} - \int \frac{d^4L}{(2\pi)^4}\frac{d^4Q}{(2\pi)^4} \frac{1}{R_L L^2 (Q+K)^2}\nonumber\\
&&\qquad\qquad\qquad\qquad\qquad\qquad\qquad\qquad\quad  \times \bigg(\frac{1}{L^2 Q^2} - \frac{1}{Q^2 (Q+L)^2} - \frac{1}{L^2 (Q+L)^2} \bigg)  \bigg\}.\qquad
\end{eqnarray}

\noindent
The angular integrations in this expression can be made with the help of the Chebyshev polynomials. After some transformations the result can be presented in the form

\begin{eqnarray}
&&\hspace*{-5mm} q(0) = \frac{1}{(8\pi^2)^2}\, \lim\limits_{K\to 0}\bigg\{- \ln\frac{\Lambda}{K}
+ \int\limits_K^\infty \frac{dQ}{Q^3} \int\limits_0^K \frac{dL\,L}{R_L} + \int\limits_K^\infty \frac{dL\,L}{R_L} \int\limits_L^\infty \frac{dQ}{Q^3} + \int\limits_K^\infty \frac{dL}{R_L L^3} \int\limits_K^L dQ\,Q
\nonumber\\
&&\hspace*{-5mm} + \int\limits_0^K \frac{dL\, L}{R_L K^2} \int\limits_L^K \frac{dQ}{Q} + \int\limits_0^K \frac{dQ\, Q^3}{K^2} \int\limits_Q^K \frac{dL}{R_L L^3} + \int\limits_K^\infty \frac{dL}{R_L L^3} \int\limits_0^K \frac{dQ\, Q^3}{K^2}\bigg\}. \qquad
\end{eqnarray}

\noindent
Some integrals in this expression can be calculated. (In some of them it is necessary to take into account that $\lim_{K\to 0} R(K^2/\Lambda^2) =1$.) Then we obtain

\begin{equation}
q(0) = \frac{1}{(8\pi^2)^2}\, \lim\limits_{K\to 0}\bigg\{- \ln\frac{\Lambda}{K} + \frac{5}{8} + \int\limits_K^\infty \frac{dL}{R_L L} - \frac{K^2}{4}\int\limits_K^\infty \frac{dL}{R_L L^3}
\bigg\}.
\end{equation}

\noindent
In the limit $K\to 0$ the last integral can be taken after the substitution $L=xK$,

\begin{equation}\label{Last_Integral}
\lim\limits_{K\to 0}\int\limits_K^\infty \frac{dL\, K^2}{R_L L^3} = \lim\limits_{K\to 0} \int\limits_1^\infty dx \frac{1}{x^3 R(x^2 K^2/\Lambda^2)} = \int\limits_1^\infty \frac{dx}{x^3} = \frac{1}{2}.
\end{equation}

\noindent
Also we note that the constant $A_1$ can equivalently be rewritten in the form

\begin{equation}\label{A1_Equivalent}
A_1 =  \lim\limits_{\varepsilon\to 0} \int\limits_\varepsilon^\infty dx\, \ln x\, \frac{d}{dx}\Big(\frac{1}{R(x)}\Big) = \lim\limits_{\varepsilon\to 0}\bigg\{\frac{\ln x}{R(x)}\bigg|_\varepsilon^\infty - \int\limits_\varepsilon^\infty \frac{dx}{xR(x)}\bigg\} = \lim\limits_{K\to 0} \bigg\{2 \ln \frac{\Lambda}{K} - 2\int\limits_{K}^\infty \frac{dL}{R_L L}\bigg\},
\end{equation}

\noindent
where we made the change of the integration variable $x=L^2/\Lambda^2$ and took $\varepsilon = K^2/\Lambda^2$. Using Eqs. (\ref{Last_Integral}) and (\ref{A1_Equivalent}) we obtain

\begin{eqnarray}
q(0) = \frac{1}{(8\pi^2)^2}\Big( \frac{1}{2} -\frac{A_1}{2} \Big),
\end{eqnarray}

\noindent
so that

\begin{equation}
I_{1,2,2} = \frac{N_f \alpha^3}{\pi^3} \Big( \frac{1}{2} -\frac{A_1}{2} \Big).
\end{equation}

\noindent
Therefore, the considered contribution to the anomalous dimension of the matter superfields takes the form

\begin{equation}\label{I12_Result}
I_{1,2}= - \Big(\ln\frac{\Lambda}{\mu} + b_{2,0}\Big) I_{1,2,1} + I_{1,2,2} + I_{1,2,3} = - \Big(\ln\frac{\Lambda}{\mu} + b_{2,0}\Big) \frac{N_f \alpha^3}{\pi^3} + \frac{N_f \alpha^3}{2\pi^3}.
\end{equation}

\subsection{Integrals quadratic in $N_f$}
\hspace*{\parindent}

Now, let us calculate the remaining contribution to the three-loop anomalous dimension of the matter superfields which is proportional to $(N_f)^2$. It comes from superdiagrams containing two insertions of the one-loop polarization operator. Also terms proportional to $(N_f)^2$ (which certainly should be taken into account) come from the renormalization of the coupling constant in the one-loop superdiagrams and in the two-loop superdiagrams proportional to $N_f$. The sum of all these contributions is given by the expression

\begin{eqnarray}
&& I_2 \equiv -8e^6 (N_f)^2 \frac{d}{d\ln\Lambda} \int \frac{d^4K}{(2\pi)^4} \frac{1}{R_K^3 K^4} \bigg\{\int\frac{d^4Q}{(2\pi)^4}\bigg(\frac{1}{Q^2 (Q+K)^2} \nonumber\\
&&\qquad\qquad\qquad\qquad\qquad - \frac{1}{(Q^2+M^2)((Q+K)^2+M^2)}\bigg) - \frac{1}{8\pi^2}R_K \Big(\ln\frac{\Lambda}{\mu} + b_{1,0}\Big) \bigg\}^2.\qquad
\end{eqnarray}

\noindent
It is convenient to rewrite it in terms of the function $p(K)$ defined by Eq. (\ref{P_Definition}). After some transformation we present the considered part of the anomalous dimension in the form

\begin{eqnarray}
&&\hspace*{-5mm} I_2 = \frac{e^6}{8\pi^4} (N_f)^2 \int \frac{d^4K}{(2\pi)^4} \frac{1}{K^4} \bigg\{- \Big(\ln\frac{\Lambda}{\mu} + b_{1,0}\Big)^2 \frac{d}{d\ln\Lambda}\Big(\frac{1}{R_K}\Big)
+ 2 \Big(\ln\frac{\Lambda}{\mu} + b_{1,0}\Big) \ln\frac{\Lambda}{K}\, \nonumber\\
&&\hspace*{-5mm} \times \frac{d}{d\ln\Lambda}\Big(\frac{1}{R_K}\Big)
+ 16\pi^2 \Big(\ln\frac{\Lambda}{\mu} + b_{1,0}\Big) \frac{d}{d\ln\Lambda}\Big(\frac{p(K)}{R_K^2}\Big)
- \ln^2 \frac{\Lambda}{K}\, \frac{d}{d\ln\Lambda}\Big(\frac{1}{R_K}\Big)
- 16\pi^2 \ln\frac{\Lambda}{K}
\nonumber\\
&&\hspace*{-5mm} \times \frac{d}{d\ln\Lambda}\Big(\frac{p(K)}{R_K^2}\Big)
- (8\pi^2)^2 \frac{d}{d\ln\Lambda}\Big(\frac{p^2(K)}{R_K^3}\Big) \bigg\}.
\end{eqnarray}

\noindent
Some integrals entering this expression can easily be calculated using the technique described above,

\begin{eqnarray}
&& I_{2,1} \equiv \frac{e^6}{8\pi^4} (N_f)^2 \int \frac{d^4K}{(2\pi)^4} \frac{1}{K^4} \frac{d}{d\ln\Lambda}\Big(\frac{1}{R_K}\Big) = \frac{\alpha^3 (N_f)^2}{\pi^3};\\
&& I_{2,2} \equiv \frac{e^6}{4\pi^4} (N_f)^2 \int \frac{d^4K}{(2\pi)^4} \frac{1}{K^4} \ln\frac{\Lambda}{K}\,\frac{d}{d\ln\Lambda}\Big(\frac{1}{R_K}\Big) = \frac{\alpha^3 (N_f)^2}{\pi^3} A_1;\\
&& I_{2,3} \equiv \frac{2e^6}{\pi^2} (N_f)^2 \int \frac{d^4K}{(2\pi)^4} \frac{1}{K^4} \frac{d}{d\ln\Lambda}\Big(\frac{p(K)}{R_K^2}\Big)\nonumber\\
&& \qquad\qquad\qquad\qquad\quad\  = \frac{2\alpha^3 (N_f)^2}{\pi^3} 8\pi^2 p(0) = \frac{2\alpha^3 (N_f)^2}{\pi^3} (\ln a +1); \qquad\\
&& I_{2,4} \equiv 8e^6 (N_f)^2 \int \frac{d^4K}{(2\pi)^4} \frac{1}{K^4} \frac{d}{d\ln\Lambda}\Big(\frac{p^2(K)}{R_K^3}\Big) \nonumber\\
&& \qquad\qquad\qquad\quad\ \ = \frac{\alpha^3 (N_f)^2}{\pi^3} \left(8\pi^2 p(0)\right)^2 = \frac{\alpha^3 (N_f)^2}{\pi^3} (\ln a +1)^2.\qquad
\end{eqnarray}

\noindent
One more integral is reduced to the constant $A_2$ defined by Eq. (\ref{Constants_Definition}) after calculating the angular integral and making the substitution $x=K^2/\Lambda^2$,

\begin{eqnarray}
&& I_{2,5} \equiv \frac{e^6}{8\pi^4} (N_f)^2 \int \frac{d^4K}{(2\pi)^4} \frac{1}{K^4}\, \ln^2 \frac{\Lambda}{K}\, \frac{d}{d\ln\Lambda}\Big(\frac{1}{R_K}\Big)\nonumber\\
&&\qquad\qquad\qquad\qquad\qquad = -\frac{\alpha^3}{4\pi^3} (N_f)^2 \int\limits_0^\infty dx\, \ln^2x\, \frac{d}{dx}\Big(\frac{1}{R(x)}\Big) = -\frac{\alpha^3}{4\pi^3} (N_f)^2 A_2.\qquad
\end{eqnarray}

\noindent
For the regulator function $R(x) = 1+x^n$ the integral which determines the constant $A_2$ can be calculated, see Eq. (\ref{ABCD1_Values}).

In the remaining integral we also make the same substitution, after which it is reduced to the expression

\begin{eqnarray}
&& I_{2,6} \equiv \frac{2 e^6}{\pi^2} (N_f)^2 \int \frac{d^4K}{(2\pi)^4} \frac{1}{K^4}\, \ln\frac{\Lambda}{K}\, \frac{d}{d\ln\Lambda}\Big(\frac{p(K)}{R_K^2}\Big) = \frac{\alpha^3}{\pi^3} (N_f)^2 \int\limits_0^\infty dx\, \ln x\, \nonumber\\
&& \times \frac{d}{dx}\bigg\{\frac{1}{R^2(x)}\bigg[ -\frac{1}{2}\Big(1-R(x)\Big) \ln x + \ln a + \sqrt{1+\frac{4a^2}{x}}\, \mbox{arctanh}\,\sqrt{\frac{x}{x+4a^2}}\,\bigg] \bigg\}. \qquad
\end{eqnarray}

\noindent
It is reasonable to extract a part proportional to $\ln a$, so that the result takes the form

\begin{equation}
I_{2,6} = \frac{\alpha^3}{\pi^3} (N_f)^2 \Big(D_1 \ln a + D_2 \Big),
\end{equation}

\noindent
where the constants $D_1$ and $D_2$ are defined by Eq. (\ref{Constants_Definition}). For the regulator $R(x)=1+x^n$ the constant $D_1$ can be found analytically and is given by Eq. (\ref{ABCD1_Values}). Unfortunately, even for this regulator we did not manage to find an analytic expression for the constant $D_2$. However, this constant can easily be calculated numerically. For $a=1$ and the regulators $R(x)=1+x$ and $R(x)=1+x^2$ the values of this constant are given by Eq. (\ref{D2_Values}).

Using the expressions for the above integrals we obtain the result for a part of the three-loop anomalous dimension proportional to $(N_f)^2$,

\begin{eqnarray}\label{I2_Result}
&& I_2 = - \Big(\ln\frac{\Lambda}{\mu} + b_{1,0}\Big)^2 I_{2,1} + \Big(\ln\frac{\Lambda}{\mu} + b_{1,0}\Big)\Big(I_{2,2} + I_{2,3}\Big) - I_{2,4} - I_{2,5} - I_{2,6} \qquad\nonumber\\
&& = - \frac{\alpha^3 (N_f)^2}{\pi^3}\Big(\ln\frac{\Lambda}{\mu} + b_{1,0}\Big)^2 + \frac{2\alpha^3 (N_f)^2}{\pi^3} \Big(\ln\frac{\Lambda}{\mu} + b_{1,0}\Big)\Big(\ln a +1 +\frac{A_1}{2}\Big)\nonumber\\
&& -  \frac{\alpha^3 (N_f)^2}{\pi^3}\Big((\ln a+1)^2 - \frac{A_2}{4} + D_1\ln a + D_2\Big).
\end{eqnarray}

\subsection{The result}
\hspace*{\parindent}

Summing up the two-loop anomalous dimension (see, e.g., \cite{Aleshin:2020gec}) and the three-loop contributions (\ref{I0_Result}), (\ref{I11_Result}), (\ref{I12_Result}), and (\ref{I2_Result}) we obtain the expression for the three-loop anomalous dimension of the matter superfields defined in terms of the bare coupling constant,

\begin{eqnarray}
&&\hspace*{-7mm} \gamma(\alpha_0) = - \frac{\alpha}{\pi} + \frac{\alpha^2}{2\pi^2} + \frac{\alpha^2 N_f}{\pi^2}\Big(\ln a + 1 + \frac{A_1}{2}\Big) - \frac{\alpha^2 N_f}{\pi^2}\Big(\ln\frac{\Lambda}{\mu} + b_{1,0}\Big) - \frac{\alpha^3}{2\pi^3} -\frac{\alpha^3 N_f}{\pi^3}\Big(\ln a + \frac{3}{4} \nonumber\\
&&\hspace*{-7mm}  + C\Big) + \frac{\alpha^3 N_f}{\pi^3} \Big(\ln\frac{\Lambda}{\mu} + b_{1,0}\Big)
- \frac{\alpha^3 N_f}{\pi^3}\Big(\ln\frac{\Lambda}{\mu} + b_{2,0}\Big)
- \frac{\alpha^3 (N_f)^2}{\pi^3}\Big((\ln a +1)^2 - \frac{A_2}{4} + D_1 \ln a + D_2\Big)\nonumber\\
&&\hspace*{-7mm}  + \frac{2\alpha^3(N_f)^2}{\pi^3}\Big(\ln\frac{\Lambda}{\mu} + b_{1,0}\Big) \Big(\ln a +1 +\frac{A_1}{2}\Big)
- \frac{\alpha^3 (N_f)^2}{\pi^3}\Big(\ln\frac{\Lambda}{\mu} + b_{1,0}\Big)^2 + O(\alpha^4).
\end{eqnarray}

\noindent
Note that the right hand side of this equation is written in terms of the renormalized coupling constant $\alpha$. Certainly, it is necessary to express it in terms of the bare coupling constant $\alpha_0$ with the help of Eq. (\ref{Relation_Between_Couplings}). After this all terms containing $\ln\Lambda/\mu$ should cancel each other. This fact can be considered as test of the calculation correctness. This is really so, and the final result for the three-loop anomalous dimension defined in terms of the bare coupling constant is given by the expression (\ref{Gamma_3Loop_Bare}), which is also presented here for completeness,

\begin{eqnarray}
&& \gamma(\alpha_0) = - \frac{\alpha_0}{\pi} + \frac{\alpha_0^2}{2\pi^2} + \frac{\alpha_0^2 N_f}{\pi^2}\Big(\ln a + 1 + \frac{A_1}{2}\Big) - \frac{\alpha_0^3}{2\pi^3} - \frac{\alpha_0^3 N_f}{\pi^3}\Big(\ln a + \frac{3}{4} + C\Big)\nonumber\\
&& - \frac{\alpha_0^3(N_f)^2}{\pi^3} \Big((\ln a + 1)^2 -\frac{A_2}{4} + D_1 \ln a + D_2\Big) + O(\alpha_0^4).
\end{eqnarray}

\noindent
As it should be \cite{Kataev:2013eta}, this expression does not depend on the finite constants $b_{1,0}$ and $b_{2,0}$ which (partially) determine the subtraction scheme in the lowest orders.

\end{document}